\documentclass[12pt,preprint]{aastex62}
\def\lea{\mathrel{<\kern-1.0em\lower0.9ex\hbox{$\sim$}}}
\def\gea{\mathrel{>\kern-1.0em\lower0.9ex\hbox{$\sim$}}}

\shorttitle{UVOT Stars: Open Clusters}
\shortauthors{Siegel et al.}

\begin{document}

\title{The {\it Swift\/} UVOT Stars Survey. III. Photometry and Color-Magnitude Diagrams of 103 Galactic Open Clusters}

\author{Michael H. Siegel}
\affiliation{Pennsylvania State University, Department of Astronomy \\
525 Davey Laboratory \\
University Park, PA, 16802}

\author{Samuel J. LaPorte}
\affiliation{Pennsylvania State University, Department of Astronomy \\
525 Davey Laboratory \\
University Park, PA, 16802}

\author{Blair L. Porterfield}
\affiliation{Pennsylvania State University, Department of Astronomy \\
525 Davey Laboratory \\
University Park, PA, 16802}
\affiliation{Space Telescope Science Institute \\
3700 San Martin Drive \\
Baltimore, MD, 21218}

\author{Lea M. Z. Hagen}
\affiliation{Pennsylvania State University, Department of Astronomy \\
525 Davey Laboratory \\
University Park, PA, 16802}
\affiliation{Space Telescope Science Institute \\
3700 San Martin Drive \\
Baltimore, MD, 21218}
\affiliation{Institute for Gravitation and the Cosmos \\
The Pennsylvania State University \\
University Park, PA 16802}

\author{Caryl A. Gronwall}
\affiliation{Pennsylvania State University, Department of Astronomy \\
525 Davey Laboratory \\
University Park, PA, 16802}
\affiliation{Institute for Gravitation and the Cosmos \\
The Pennsylvania State University \\
University Park, PA 16802}

\begin{abstract}
As part of the Swift/UVOT Stars Survey, we present near-ultraviolet (3000-1700 \AA ) point-source photometry for 103 Galactic open clusters.  These data, taken over the span
of the mission, provide a unique and unprecedented set of near-ultraviolet point-source photometry on simple stellar populations.
After applying membership analysis fueled mostly by GAIA DR2 proper motions, we find that 49 of these 103 have clear precise
CMDs amenable to investigation.  
We compare the CMDs to theoretical isochrones and find good agreement between the theoretical isochrones and the
CMDs. The exceptions are the fainter parts of the main sequence and the red giant branch in the $uvw2-uvw1$ CMDs, which is most likely due
either
to the difficulty of correcting for the red leak in the $uvw2$ filter or limitations in our understanding of UV opacities for
cool stars.  For the most part, our derived cluster parameters -- age, distance and reddening -- agree with the consensus literature but we find a few clusters
that warrant substantial revision from literature values, notably NGC~2304,  NGC~2343, NGC~2360, NGC~2396, NGC~2428, NGC~2509, NGC~2533, NGC~2571, NGC~2818, Collinder~220 and
NGC~6939. A number of these are clusters
in the third Galactic quadrant where previous studies have mistaken the disk sequence for the cluster. However, the GAIA DR2 proper motions
clearly favor a different sequence.
A number of clusters also show white dwarf and blue straggler sequences.  We confirm the presence of extended main sequence turnoffs in
NGC~2360 and NGC~2818 and show hints of it in a number of other clusters which may warrant future spectroscopic study. 
Most of the clusters in the study have low extinction and the rest are well fit by a ``Milky-Way-like"
extinction law.  However, Collinder~220 hints at a possible ``LMC-like" extinction law.
We finally provide a comprehensive point-source catalog to the community as a tool for future investigation.
\end{abstract}


\section{Introduction}
\label{s:intro}

Open clusters are gravitationally bound collection of $10^2$-$10^4$ stars born in the same star formation
event.  They are, by far, the most abundant family of clusters in the Milky Way with over 1500 cluster or cluster candidates
having been cataloged (Dias et al. 2002)
Galactic open clusters are predominantly young and metal-rich -- most of the older clusters having
long since dissolved into the Galactic field.  However, a number of old and metal-poor clusters have been detected and serve
as ``fossils" for studying the early Galaxy.

As compact collections of stars of similar age, abundance, distance and reddening, individual 
star clusters present a snapshot of stellar evolution.  Studying large samples of clusters
allows us to piece together the narrative of stellar evolution and constrain the photometric properties
of stars as functions of mass, age and chemistry.
Much of our understanding of stellar evolution and photometry comes from the study of Galactic open and globular
star clusters.  Cluster of all ages and chemistries are known to host stars that are bright in 
the ultraviolet (UV), as shown in Siegel et al. (2014, hereafter Paper I).  These UV-bright stars in the nearby universe
have counterparts in the UV-bright stars seen in nearby galaxies and are, ultimately, parallels to the stars contributing
to the UV light of distant unresolved stellar populations.  Any understanding
of luminous hot UV-bright stars and stellar populations dominated by them will therefore have to lay its
foundation on the study of nearby star clusters.  They are not only the best repositories
in which to find such stars, but having found them, the properties can be immediately connected with stellar populations of known age, distance, reddening
and chemistry.

The study of the UV properties of star clusters is still in its early stages, as detailed in Paper I.  But there a number of recent results
that indicate this is a fertile field of study.  For example, studies of open clusters in the Magellanic Clouds have shown complex structure in the 
main sequence turnoff (MSTO) -- an extended MSTO (eMSTO) and split MSTO's (Mackey \& Broby Nielsen 2007; Milone et al. 2009; Goudfrooij et al. 2011).  This
split has also recently been identified in a number of Galactic clusters (Marino et al. 2018. Cordoni et al. 2018)
This fine structure
is thought to the be the result of stellar rotation, which can affect the evolutionary lifetimes of stars, and may show up more clearly in the UV.
A recent comprehensive study of M67 (Sindhu et al. 2018) identified numerous blue straggler stars (BSS), white dwarf binaries and stars that had excess
far-UV emission consistent with either unresolved binarism or chromospheric activity. 

In addition to their utility for studying stellar evolution and the properties of rare phases of stellar
evolution, Galactic open clusters can be used as the building blocks for studying larger astrophysical questions.
Open clusters, for example, can be used to probe the chemodynamical properties of the Galactic
disc (see, e.g., Friel 1995, Twarog et al. 1997, Yong et al. 2005, 2012; Frinchaboy \& Majewski 2008).
Old open clusters, with ages greater than 1 Gyr, are particularly suited to tracing the chemical evolution of the disk through their age-metallicity
relationship (AMR).  Indeed, the similarity between the AMR of a small group of open clusters and those of merging dwarf
galaxy systems has been key to confirming past Galactic mergers (Frinchaboy et al. 2006).  Open clusters may be particularly useful
for probing the UV extinction properties of nearby dust.  As we have noted before (Paper I, Hagen et al. 2017, 2019), the UV extinction curve is uncertain, both
in slope and in the strength of the ``blue bump" at 2175 \AA\
(Stecher 1965; Seaton 1979; Gordon et al. 2003).  Open clusters can serve as probes of dust properties on small and large scales.

Of particular interest to recent scientific investigations is the utility of clusters for studying the properties of
complex stellar populations. Open clusters are nearby, meaning they can be probed on a
star-by-star basis to determine age, abundance, foreground reddening and distance.  However, they can also be studied through their {\it integrated light}.
By combining these two measures,
one can calibrate the study of more distant stellar populations that cannot be resolved on a star-by-star basis but can {\it only} be studied though
the properties of their integrated light (see an illustration of this in the context of the Magellanic Clouds in Searle et al. 1980).
Studies of the integrated light of unresolved stellar population have proven extraordinary useful for untangling the star-formation histories and dust
extinction properties of both nearby and distance galaxies (see, e.g., Hoversten et al. 2011, Calzetti et al. 2015; Hagen et al. 2017, 2019).
However, these studies have a limitation: the UV portion of the spectral-energy distribution is not as well constrained as the optical and infrared.
The lack of empirical
information needed to calibrate synthetic spectra of unresolved stellar populations was specifically cited by Bruzual (2009) as a limitation on the utility of the models.

Despite the tremendous efforts made by investigators, most of the Milky Way's open clusters have not been studied in great detail at {\it any} wavelength.
The sheer number of open clusters in the Galaxy precluded most from being targeted for individual study. Indeed, many of the parameters used by the WEBDA online database
of open clusters\footnote{\url{http://www.univie.ac.at/webda/}} come from global optical-infrared (OIR) surveys of hundreds of clusters such as that of Kharchenko et al. (2013 and 2016, hereafter 
K13 and K16, respectively).  But there are
additional issues that complicate the study of open clusters, most notably their preferred location within the Galactic midplane, with concomitant heavy foreground
reddening and field-star contamination.

This deficit of individual study is particularly acute in the UV where very few clusters have been studied in detail (see discussion in Paper I).
This lacuna is the natural apotheosis of the factors that limit their study in the optical:  their 
location within the Galactic midplane makes study difficult either due to crowding (most previous UV missions had coarse spatial resolution), field
star contamination, foreground extinction (which is much stronger in the UV) or simple brightness constraints that preclude missions like GALEX
from studying objects in the Galactic midplane.

As a result of this, the spectral synthesis models used to study distant unresolved stellar populations are still not well-constrained in the UV.  And this is particularly
problematic because young stellar populations -- those most visible over the long distances in extragalactic astronomy and cosmology -- emit a large fraction of their
rest-frame light in the UV.  {\it The lack of thorough, systematic and empirical study of young
UV-bright stellar populations is a significant and limiting
problem for extant and planned ultraviolet surveys of the Milky Way, the Magellanic Clouds, Local Group objects, nearby galaxies and
distant galaxies}.  Understanding the star-by-star and integral photometry of all these distant and extragalactic
systems -- from the nearby to the deep universe -- is critically dependent on studies of nearby open clusters.

In this contribution, we address this deficit in our knowledge of Galactic open clusters with a survey of 103 open clusters observed in the near-ultraviolet (NUV) with
the Neil Gehrels Swift Mission's Ultraviolet-Optical Telescope (UVOT).  By combining Swift's wide-field moderate resolution photometry with precise
astrometry from the recent GAIA DR2 catalog, we are able to study these clusters in unprecedented detail, testing the utility of theoretical isochrones
and either confirming or revising the literature parameters for these clusters.

We present the observational details of the program in \S \ref{s:obsred}.  We then detail the limitations on our
analysis produced by saturation of the brightest stars and
contamination from the Galactic disk in \S \ref{s:analysis}.  Using some
radial velocities but mostly GAIA DR2 proper motions to establish membership, we are able to identify 49 clusters that have clear and
distinct color-magnitude sequences, many of which have never been the target of detailed study.  We present these color-magnitude
diagrams (CMDs) along with comparison to theoretical isochrones, with which we measure or revise their fundamental properties.  The resultant
test of the theoretical isochrones shows excellent agreement, indicating that the models perform well in the UV.  We show that the isochrones
have a tendency to not match the photometry of cooler stars -- red giant branch (RGB) and late-type main sequence (MS) stars.
We finally look ahead to other uses for this dataset, which we will provide to the community.

\section{Observations and Data Reductions}
\label{s:obsred}

\subsection{UVOT Data}
\label{ss:uvot}

UVOT is a modified Richey-Chretien 30 cm telescope that has a 
wide (17' $\times$ 17') field of view and a microchannel plate intensified CCD operating in photon
counting mode (see details in Roming et al. 2000, 2004, 2005) aboard the Neil Gehrels Swift Gamma Ray Burst mission (Gehrels et al. 2004).
The instrument is equipped with a filter wheel that includes a clear white filter, $u$, $b$ and $v$ optical filters, 
$uvw1$, $uvm2$ and $uvw2$ ultraviolet filters, a magnifier, two grisms and a blocked filter.  Although its primary
mission is to measure the optical/ultraviolet afterglows
of gamma ray bursts, the wide field, 2\farcs3 resolution, broad wavelength range (1700-8000 \AA)
and ability to observe simultaneously with {\it Swift}'s X-Ray Telescope (XRT; Burrows et al. 2005)
allow a broad range of science, including the study of hot or highly energetic stars.  It is particularly ideal, in the context of hot stars,
for studying nearby star clusters.  Its wide field
can enclose most nearby open clusters in a single pointing and its resolution allows measurement of stars almost to the center of the most crowded
fields. The NUV filters -- with effective wavelengths of 2600, 2246 and 1928 \AA\ for $uvw1$, $uvm2$ and $uvw2$, respectively, provide good coverage
of the NUV wavelength range and sensitivity to the 2175 \AA\ blue bump.

The open clusters surveyed by UVOT are drawn from a variety of programs.  The bulk were observed as part of M. Siegel's approved Cycle 12 GI and team fill-in programs,
observed optimally for 6 ks in UVOT's three NUV filters (although, in practice, the observing time tended to be closer to 5 ks).  Others were observed as calibration
targets or coincidentally with other studies.  We have only included clusters in the sample if they were observed for at least 0.8 ks in two of UVOT's
three NUV filters and where the field center was within 5' of the nominal cluster center, although some exceptions have been made to expand
the sample.
The clusters in our GI and fill-in program were generally selected to be comparable to the UVOT field-of-view, with low reddening ($E (B-V) < 0.5$) and locations
away from Galactic Center region, which is the most heavily oversubscribed part of the sky for Swift observations.

The list of 103 clusters is given in Table \ref{t:swiftphot}.  
The table lists the exposure time in UVOT's three NUV passbands as well
as its broadband $u$ passband.
All exposure times are given as the maximum exposure time on the combined image, which removes deadtime losses.
Most of the cluster images were created by combining any image within 7\farcm5 of
the cluster center.  However, some of the first images processed in the program used a much larger search radius in an effort to provide
data on the clusters with the richest history of prior observations.  All clusters were processed
and photometered through the pipeline described in Paper I, updated to the most recent calibrations (Breeveld et al. 2011).

While this particular contribution focuses on the clusters with the cleanest color-magnitude diagrams, we have created photometric catalogs
for all 103 clusters.  The photometric catalogs include calibrated PSF photometry in all available passbands, including broadband $b$ and $v$
when available.  They have been filtered of objects with DAOPHOT (Stetson 1987, 1994)
structural parameters of $|SHARP| > 0.5$ to remove blends and detections near bright saturated stars (see Paper I for more information).  The catalogs are currently available
on request but will eventually be made available through the MAST archive.

Of the clusters in the sample, all but one have $uvw1$ data, all but one
have $uvm2$ data and all but three have $uvw2$ data.  Only two clusters -- Hogg~16 and Westerlund~2 -- lack $uvm2$ and $uvw1$ data.
Our sample includes few clusters with very young ages ($t < 100$ Myr).  The reason
for this is that very young clusters tend to be either far too bright or far too enshrouded in dust for UVOT study.

\begin{deluxetable}{ccccccc}
\tablewidth{0 pt}
\tabletypesize{\footnotesize}
\tablecaption{Swift/UVOT Observations of Open Clusters\label{t:swiftphot}}
\tablehead{
\colhead{Cluster} &
\colhead{u Exp} &
\colhead{uvw1 Exp} &
\colhead{uvm2 Exp} &
\colhead{uvw2 Exp} &
\colhead{Observation}\\
\colhead{} &
\colhead{Time (ks)} &
\colhead{Time (ks)} &
\colhead{Time (ks)} &
\colhead{Time (ks)} &
\colhead{Dates}}
\startdata
\hline
         Blanco~1    &        0 &      3378 &      2461 &      3071 &  2010-11-10 -- 2011-01-25   \\
\hline
          NGC~188    &     1195 &      2842 &      3111 &      4555 &                2005-06-29   \\
          \nodata    &  \nodata &   \nodata &   \nodata &   \nodata &                2005-10-13   \\
          \nodata    &  \nodata &   \nodata &   \nodata &   \nodata &  2007-10-30 -- 2007-12-06   \\
\hline
          NGC~752    &        0 &      798 &      1032 &      1076 &  2011-07-31 -- 2011-09-06   \\  
\hline
          IC~1805    &        0 &       862 &       926 &       926 &                2008-04-06   \\
\hline
         NGC~1039 (M~34)    &        0 &      2137 &      2206 &      2206 &                2011-03-08   \\
          \nodata    &  \nodata &   \nodata &   \nodata &   \nodata &  2011-07-10 -- 2011-07-14   \\
\hline
         NGC~1896    &        0 &      1717 &      1614 &      1614 &                2013-10-27   \\
\hline
         NGC~1996    &        0 &      3490 &      3428 &      3507 &  2013-10-20 -- 2013-10-26   \\
\hline
\hline
\enddata
\footnotesize{This table is available in its entirety in machine-readable and Virtual Observatory (VO) forms.}
\end{deluxetable}

\section{Analysis and Results}
\label{s:analysis}

\subsection{Bright Star Issues}

Most of our cluster fields are close to or within the Galactic midplane. They also tend to be young ($t < 1$ Gyr) and nearby ($m-M < 14$).
As a result, a number of
cluster fields have numerous bright saturated stars.  ``Saturated",
in the case of a photon-counting instrument like UVOT, does not mean ``saturated" in the same sense as a CCD detector.  What this means is
that the incident count rate exceeds 372 sec$^{-1}$, the limit
at which UVOT can no longer count incident photons (corresponding to AB magnitudes of 12.52, 12.11 and 12.68 in the $uvw1$, $uvm2$ and $uvw2$ filter, respectively).
Thus, only crude lower limits can be applied to the brightness of these stars.
Even before reaching such a high level, however, the coincidence loss becomes very strong and the dead-time of the detector for incident
photons will result in dark boxes around the saturated cores of bright stars (Poole et al. 2010). This lowers the limit at which useful photometry can be measured to well
below 372 sec$^{-1}$. This is particularly noticeable
with PSF photometry as very high coincidence loss both changes the shape of the PSF for bright stars and can cause DAOPHOT to underestimate the background sky level.
In the discussions below,
the phrase ``saturated" stars will be applied to both contexts and has the general meaning of a star whose coincidence loss is
too strong for reliable PSF photometry.
While Page et al. (2017) recently outlined a method to measure photometry from saturated
stars using readout streaks, this method was not appropriate for most of our saturated stars as it would not produce the precision needed to constrain theoretical isochrones.
We have noted, within the text, several
clusters where the
bright end of the MS is saturated and we can only provide upper limits on the age of the clusters.  Two clusters that were observed for our program -- NGC~2547 and NGC~2516 -- 
are not included in the sample of 103 presented here because of bright
star contamination problems that made photometry of any kind impossible.

The other complication of bright stars is that they produce diffuse light which may affect photometry (see extended discussion in the context of M~31 in
Hagen et al. 2019).  This can
include internal reflection, donuts and read-out lines.  However, while scattered light is a problem for the photometry of extended sources, we have not found it to be a problem for {\it point sources}, which
are measured from a local sky.  Our comparison of UVOT photon-counting photometry to ground-based CCD photometry in Paper I showed that our methods produces reliable linear photometry
for all but the brightest stars.

\subsection{Cluster Membership}

While some of the clusters in our sample -- such as M~67 -- are above the Galactic plane, the vast majority are within the plane ($\mid b \mid < 10$).  This can create a problem in that the Galactic
field star population is very populous in these fields, potentially overwhelming the cluster population or confounding analysis of its color-magnitude sequence. In some fields,
the correlation between distance/reddening/age causes the field stars to vaguely resemble a MS (see, e.g., NGC~2360 in \S \ref{ss:ngc2360}).  And even in the most spartan fields, it is difficult
to identify rare phases of stellar of evolution such as blue stragglers or white dwarfs, based on photometry alone.

We approached this problem on several fronts.  Earlier proper-motion samples proved too imprecise in most cases as the cluster proper motions were within the field star distribution
and the studies lacked the precision to tease them out.  Spatial analysis -- using the parameters of K16 to select stars within the core or half-light radii, proved
effective for massive clusters but not for small ones.  Radial velocity surveys -- such as that of Mermilliod et al. (2008, hereafter MMU) proved useful but are rarely extensive
enough to cover much of the cluster and were primarily focused on bright red RGB stars.
Such stars are ideal for optical studies but are faint and poorly constrained in the UV.

Ultimately, we settled on using the proper-motion data provided in the recent GAIA Data Release 2 (Gaia Collaboration 2016, 2018).  The GAIA data are uniform, precise and have a depth
similar to that of UVOT.  Cantat-Gaudin et al. (2018, hereafter CG18) recently published a comprehensive membership analysis of 1212 clusters -- including almost all of our program clusters --
using five-dimensional phase-space information.  We have used their membership analysis for cluster membership in the UVOT data, identifying as members stars with probabilities greater than 
90\%.  For a handful of clusters, however, we found that the CG18 study only identified a small number of member stars.  For these clusters, we did our own membership analysis, as detailed
in the text.

\begin{figure}[h]
\begin{center}
\includegraphics[scale=.5,angle=270]{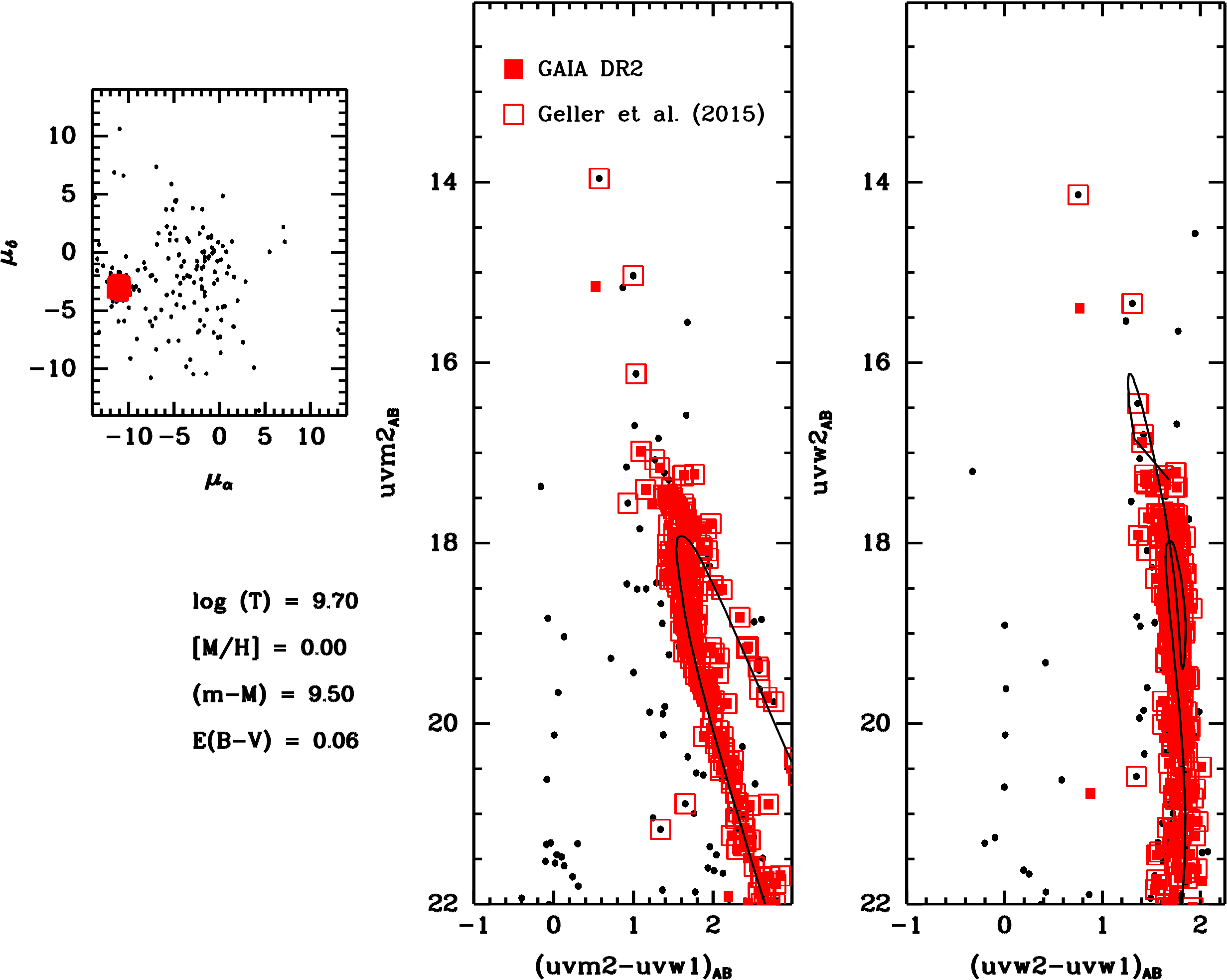}
\end{center}
\caption{GAIA DR2 vector-point (VPD) and Swift/UVOT color-magnitude diagrams of the old open cluster M~67 compared the PARSEC-COLIBRI isochrones fit in Paper I.  The solid red squares show astrometric members
selected from GAIA DR2 while the open red squares show radial-velocity members selected from Geller et al. (2015).\label{f:gaiam67}}
\end{figure}

Figure \ref{f:gaiam67} demonstrates the utility of the GAIA data in the context of M~67.  The left panel shows a vector-point diagram (VPD) of the M~67 field. The dispersed population toward the
center is the Galactic disk while the concentrated cluster of points off center is the cluster.  Almost all of the radial-velocity members from the recent survey of Geller et al. (2015)
land within the offset clump.  For the majority of the open cluster the separation between cluster and field was not quite so dramatic.
More often, the cluster proper motion was {\it within} the field star distribution. However, the GAIA DR2 parallaxes and proper motions are so precise that the cluster
members still show up a very tight clump of stars within the more dispersed Galactic background and can be easily disentangled from the field star population.

After extensive analysis, we found that 49 of our 103 target clusters could both easily be distinguished from the field star population and
showed a clear sequence in the color-magnitude diagrams.
We note that there are a number of clusters that show clear clumps of cluster stars in the GAIA data itself but fail to cross-match enough stars to the UVOT data for analysis, either because
the clusters are too faint or are too spatially dispersed. The photometry catalogs included all photometric measures for both members and non-members.

\subsection{Isochrone Fitting and Analysis}

The goal of this paper is to compare the point-source photometry of the clusters to theoretical isochrones in order to test the utility of the isochrones
(and, by extension, the atmospheric models underpinning them) in the context of young to intermediate age stellar populations.
The isochrones chosen for this exercise are the
PARSEC-COLIBRI isochrones of Marigo et al. (2017, hereafter M17).  These isochrones cover the entire range of ages and metallicities of the open cluster sample.  The
most significant revisions of this iteration are to the thermally pulsing asymptotic giant branch (TP-AGB) which is not relevant to our analysis as these stars are too cool
for good constraint with UVOT.  in fact, the detailed treatment of this phase cause the isochrones to loop around 
as the theoretical stars'
UV emission (actually, their red leak into the UV passbands, see Appendix to Paper I) waxes and wanes during their TP-AGB phases. For the figures below, the 
TP-AGB phase has been removed from the isochrones for the sake of clarity.

Isochrones were initially laid down using literature values - from studies specific to the individual cluster if possible, from K16 where no previous focused study had been made.
The isochrones were then adjusted interactively to better overlap the GAIA-selected sequences, with variation allowed in age, metallicity, reddening, distance and reddening law
(either Galactic, SMC or LMC,
based on the formulations of Pei 1992).   The main degeneracy seen in the UV isochrone fitting involved metallicity.  With all the parameters allowed to be free, the metallicity
of any cluster became unconstrained.  Any change in metallicity could be accommodated by responsive changes in distance modulus and reddening.  We therefore fixed the metallicity
of the clusters to literature spectroscopic or photometric values, where available, and solar metallicity otherwise.  For the most part, we ignored the RGB stars since very few are detected
in the NUV and those that are tend to be faint and dominated by red leak.  We do note below a few clusters such as NGC~2477 (\S \ref{ss:ngc2477})
where the RGB is prominent.

\begin{figure}[h]
\begin{center}
\includegraphics[scale=.7,angle=270]{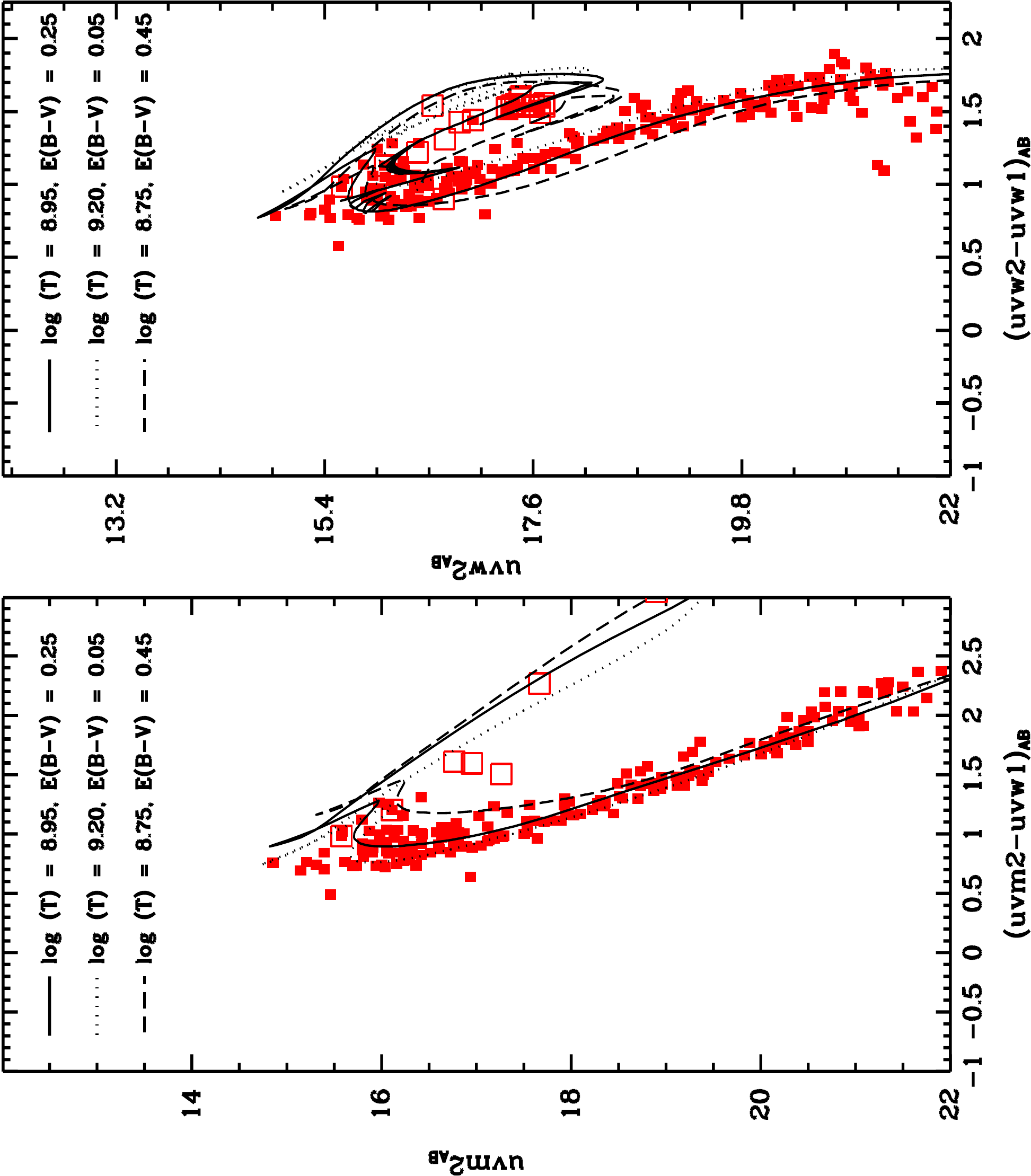}
\end{center}
\caption{A comparison of different isochrone fits to the photometry of the globular cluster NGC~2360.  The three lines are all set at a metallicity of [M/H]=-0.1 and a distance modulus
of $m-M=10.05$. However, they set at different reddening values, with age adjusted to better match the MSTO. Note that increasing or decreasing the reddening has {\it opposite} effects
in the two CMDs.  A lower (higher) reddening moves the isochrone below (above) the observed MS in the $uvm2-uvw1$ CMD and above (below) in the $uvw2-uvw1$ CMD. This allows the reddening
to be easily constrained by the observational data.
\label{f:reddemo}}
\end{figure}

Other parameters, however, were easily constrained.
The UV photometry proved particularly adept at measuring foreground reddening.  Due to the differing orientations of the
reddening vector -- mostly
blueward in the $uvm2-uvw1$ diagram and faintward in the $uvw2-uvw1$ diagram, an incorrect reddening produces a discrepancy in the location of the MS for a given
distance modulus.  With too much reddening, for example, the $uvm2-uvw1$ isochrone tends to end up above the MS while the $uvw2-uvw1$ isochrone tends to end up below (Figure \ref{f:reddemo}).  Only
the correct reddening value allows both isochrones to line up with the MS. This allowed age and distance modulus to be varied independently to match the MSTO.
We are therefore confident in our measures of foreground reddening, age and distance {\it given the assumed metallicity}.

Most of our program clusters were deliberately chosen to have low foreground reddening, which precluded any analysis of the reddening law itself
(set, by default, to the Milky Way reddening law).  However, for a few clusters, we were able to explore this issue to a modest extent (see, e.g., Collinder~220, \S \ref{ss:coll220})

As cataloged in Table \ref{t:clusparam} and in the notes on individual clusters, the isochrones performed very well. In most cases, we found that the photometry was consistent
with the predictions of theoretical isochrones set to literature values.  For a number of clusters, however, we had to substantially revise parameters to produce consistent fits.
This was particularly common among
third quadrant clusters in which the literature values sometimes described the foreground/background {\it disk} sequence, rather than the cluster.  In this region of the sky, the
reddening along the line of sight causes the ``blue wall" of disk MSTO stars to veer redward as it gets fainter, making it resemble a MS.  However,
applying the GAIA astrometric membership (as well as radial velocity memberships, when available) made it clear that this was not the cluster sequence, which was well-defined
in the GAIA astrometry.

Notes on the individual clusters are detailed below, in order of ascending Galactic longitude.  Because this is a large survey program,
we do not go into great detail on our program clusters, many of which
are worthy of a paper of their own. However, we do comment on 
discrepancies between our derived values and those found in the literature, the presence of unusual stellar types in any cluster and any potential broadening in the MSTO.

\subsection{Individual Clusters}
\label{ss:indy}

\begin{figure}[h]
\begin{center}
\includegraphics[scale=.5,angle=270]{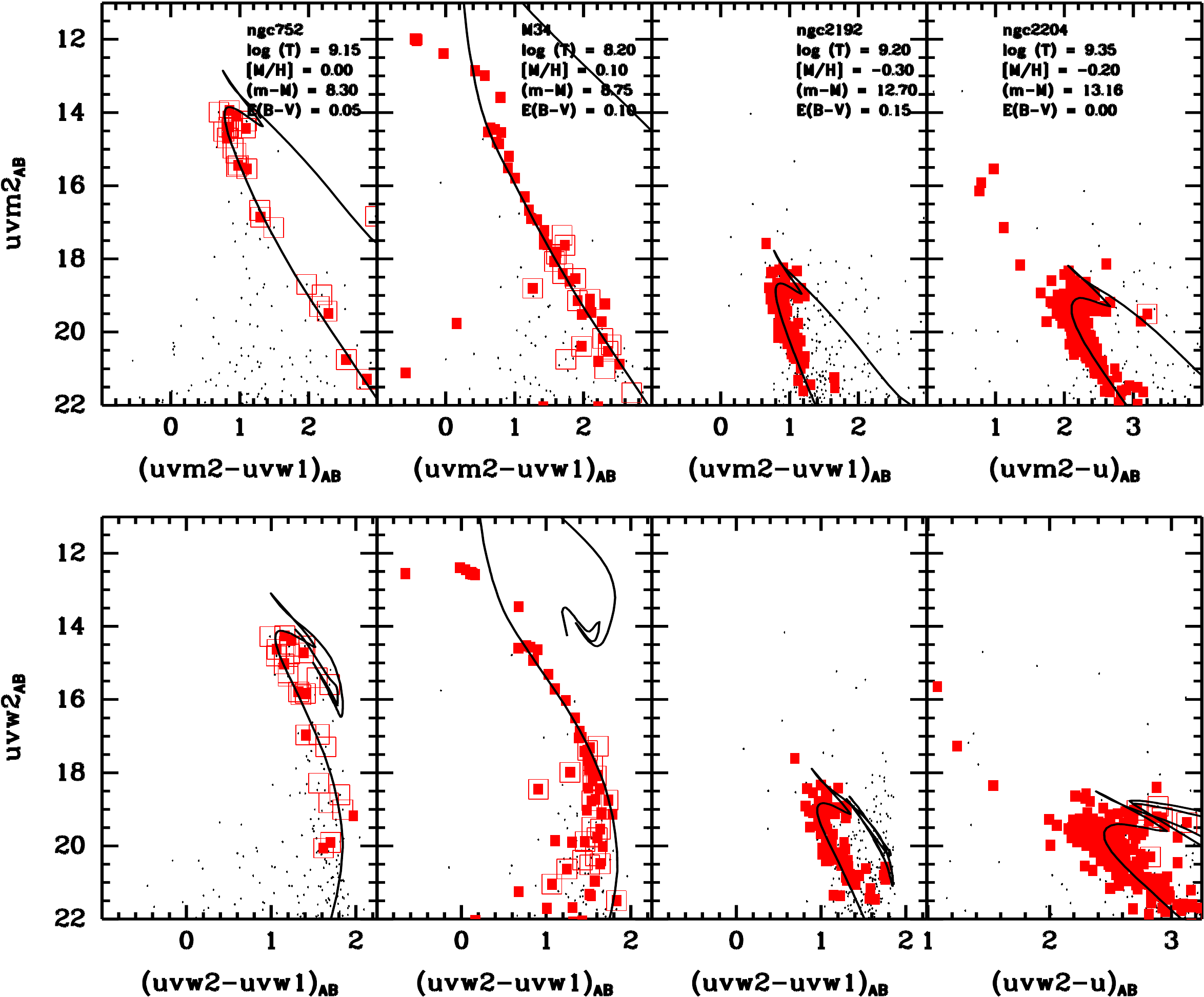}
\end{center}
\caption{Color-magnitude diagrams of the open clusters NGC~752, M~34, NGC~2192 and NGC~2204 (left to right). The solid lines are PARSEC-COLIBRI isochrones
set to the parameters in the text.  Solid red squares show astrometric members
selected from GAIA DR2 by CG18 while open squares are spectrosopic members from either MMU or sources given in the text.\label{f:CMDfig0}}
\end{figure}

\subsubsection{NGC~752}

NGC~752 is situated well below the Galactic midplane in the second Galactic quadrant.  It is very well-studied, with several cluster members established
as photometric standards and numerous thorough photometric and spectroscopic surveys in the literature.  Its age (1.4 Gyr), detailed abundance ratios and projected orbit make
NGC~752 a typical representative of the family of old disk clusters (Carraro \& Chiosi 1994; Maderak et al. 2013,
B{\"o}cek Topcu et al. 2015, Twarog et al. 2015). Its similarity to disk red giants is evidence that much of the disk population formed
from disrupted clusters (Reddy et al. 2012a,b).

The UVOT field only encloses the central regions of this dispersed cluster (K13 measure a radius of 1.4 degrees but a core radius of 3\farcm0). However, the GAIA
DR2 membership selection shows a clear MS (Figure \ref{f:CMDfig0}, first column).
We cross-identify numerous spectroscopic members catalogued by Daniel et al. (1994), all of which lie within the CG18 selection.
The NGC~752 member stars align almost perfectly with the isochrone predicted by the parameters given in Twarog et al. (2015), which is expected given how well-studied and catalogued
this cluster is.

\subsubsection{M~34}

M~34 is second quadrant cluster situated well below the Galactic midplane.  It has been extensively studied and shown to be
a solar or slightly super-solar abundance intermediate age (200 Myr) moderately-reddened ($E(B-V)\sim0.10$) cluster (Jones \& Prosser 1996; Schuler et al. 2003).  K13
measure an outer radius of 43\farcm8, indicating that M~34 fills the entire UVOT field and we are only able to image the central regions of the cluster (K13 measure a core radius of 9\farcm6).

The GAIA astrometry shows a clean separation from the field that contains the spectroscopic members listed by Meibom et al. (2011) and traces a clear MS (Figure \ref{f:CMDfig0}, second column).
Comparison to the M17 isochrones yields a distance and reddening consistent
with previous studies.  We can not constrain the age of the cluster as the upper main sequence is saturated on the UVOT exposures.  However, we can apply a lower limit to the age of
320 Myr, given the extent of the MS we detect. We also show a number of cluster member stars blueward of the primary sequence. These stars could represent the top of a faint white dwarf sequence.

The lower end of the MS differs from the isochrones in the ($uvw2-uvw1$) color-magnitude
diagram (lower panel).  As shall be seen, this discrepancy is common in clusters that are close enough for UVOT to probe the lower MS.
We should note that this is near the point at which the red leak begins to take over the isochrones as these stars
are too cool ($T_{eff} < 6300 K$) to produce significant UV light.  Indeed, the only way to make the isochrone match the data on the lower MS is to
significantly {\it increase} the reddening (which causes the isochrones to miss the upper MS in both panels).  As the reddening
increases, these faint stars become ``bluer" because the red leak flux in the $uvw1$ band is extinguished faster than the red leak flux in the $uvw2$ band. Given the known
issues with fitting UV isochrones to faint red stars (\S \ref{s:discussion}), we elected with this cluster (and other clusters demonstrating similar issues) to be guided by the
bright MS.

\subsubsection{NGC~2192}

NGC~2192 is a second quadrant cluster located above the Galactic midplane. Photometry from Park \& Lee (1999) and Tapia et al. (2010) indicate that it is old (1.3 Gyr) and metal-poor
($[M/H]\sim-0.3$), an assessment shared by the global surveys of Paunzen et al. (2010) and K16. It does not have a previous spectroscopic survey.

The astrometric selection results in a clear MS in both
CMDs (Figure \ref{f:CMDfig0}, third column).  We find that the cluster is well-fit by the parameters derived by previous investigators with a slightly older age of 1.6 Gyr, assuming the photometric
metallicity of $[M/H] = -0.3$. We identify
one potential member star brighter and bluer than the nominal MSTO.  This star is identified as a 100\% likely member by CG18 and could represent a blue straggler.
The MSTO of NGC~2192 does show some broadening beyond what would be expected from photometric errors alone. That may indicate that NGC~2192 has an extended main-sequence turnoff,
although it is a bit at the older range for clusters that would have potential eMSTOs.
It could also represent either photometric scatter in an old faint cluster or differential reddening in a moderately-reddened cluster. Our analysis
shows that differential reddening of even 0.05 magnitudes would be enough to create the relevant scatter given the high sensitivity of the NUV filters to reddening. However, there
are not enough bright stars in NGC~2192 to measure any spatial variation.  Spectroscopic study could resolve this issue.

\subsubsection{NGC~2204}

NGC~2204 is situated 16 degrees below the Galactic plane in the third quadrant. It is a slightly-metal-poor ($[M/H]=-0.20$) old (1.6 Gyr) cluster (Kassis et al. 1997; Jacobson et al. 2011). While
it has been surveyed spectroscopically, these surveys have focused on old red giant stars, most of which are faint in the UV.
Our data do not include the $uvw1$ passband but do include the redder $u$ passband.

The last column of figure \ref{f:CMDfig0} shows the CMDs compared to the M17 isochrones. The astrometric selection overlaps the member stars from the radial velocity
surveys.  Selecting stars as members based on either astrometry or radial velocity reveals a clear sequence of stars in the CMD.
Our isochrone fits to the cluster favor a slightly older age (2.3 Gyr) than previous studies with minimal reddening (although slightly larger
reddening with an SMC-like extinction law would also be consistent with the data).  Note that a NGC~2204 has a prominent
population of blue stragglers, which are all confirmed as astrometric members.  This confirms the detection of BSS by Frogel \& Twarog (1983).

\subsubsection{NGC~2243}

\begin{figure}[h]
\begin{center}
\includegraphics[scale=.5,angle=270]{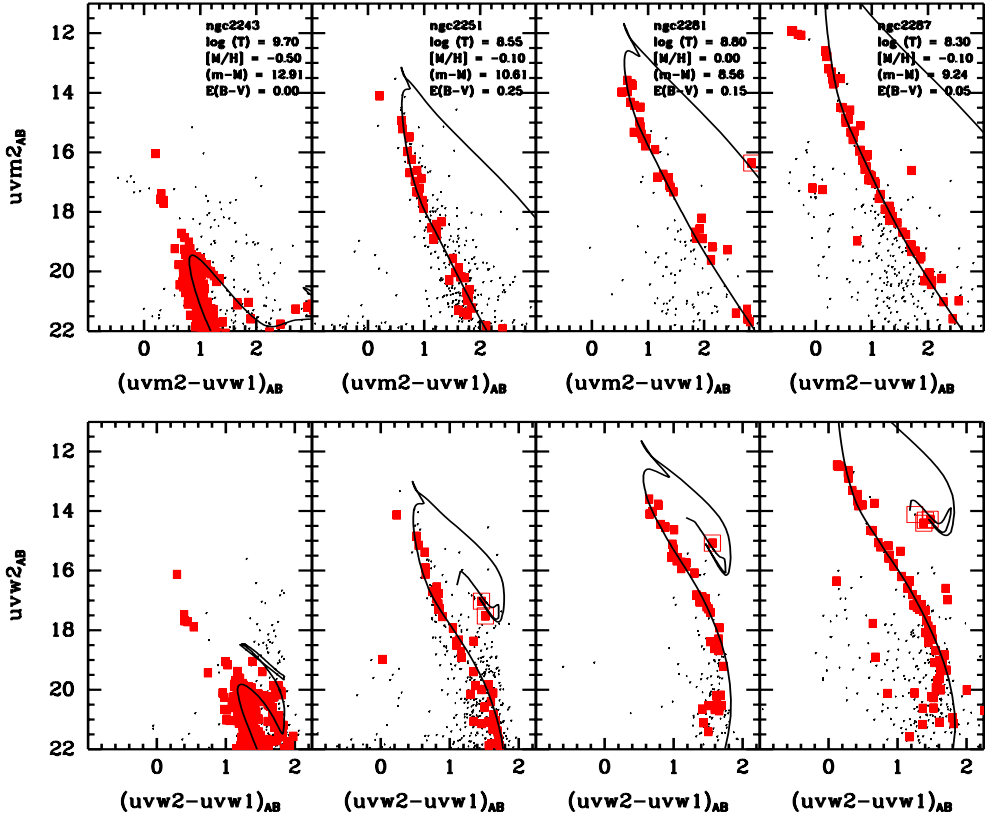}
\end{center}
\caption{Color-magnitude diagrams of the open clusters NGC~2243, NGC~2251, NGC~2281 and NGC~2287 (left to right). The solid lines are PARSEC-COLIBRI isochrones
set to the parameters in the text.  Solid red squares show astrometric members
selected from GAIA DR2 by CG18 while open squares are spectrosopic members from either MMU or sources given in the text.\label{f:CMDfig1}}
\end{figure}

NGC~2243 is a third quadrant cluster located well below the Galactic midplane.  Previous studies by Twarog et al. (1997) and Jacobson et al. (2011) have found it to be old (5 Gyr)
and metal poor ($[M/H]=-0.42$).  The radial velocity survey of MMU only measured two member stars, both of which are faint RGB stars that do not have UVOT counterparts.

Stars selected from CG18 form clean sequences in the CMDs (Figure \ref{f:CMDfig1}, first column) that are well-fit by the M17 isochrones
with parameters similar to previous studies, with slightly lower metallicity ($[M/H]=-0.5$) and minimal reddening.
We also identify a number of blue straggler stars that are astrometric members.

\subsubsection{NGC~2251}

NGC~2251 is a third-quadrant cluster located near the Galactic midplane.  Parisi et al. (2005) and Reddy et al. (2013) have studied the cluster, revealing it to be slightly
metal-poor ($[M/H]=-0.1$), intermediate age (300 Myr) and with moderate reddening ($E(B-V)=0.20$).

The astrometric selection includes one spectroscopic member from MMU.  The member stars show a clear
CMD sequence that corresponds almost exactly to the consensus literature values, with slightly higher reddening ($E(B-V)=0.25$) and age (Figure \ref{f:CMDfig1}, second column). We do note that the MSTO is near the
saturation limit but we do not find a large number of saturated stars in the image, so it likely this represents the true MSTO and thus the true age of the cluster. 

\subsubsection{NGC~2281}

NGC~2281 is positioned 16 degrees above the Galactic midplane in the second quadrant.  Studies by Glaspey (1987) and Netopil (2017) indicate that it is intermediate age (600 Myr)
 and solar metallicity.  NGC~2281
is also one of the few clusters to have been previously studied in the UV, using FUV data from GALEX (Smith 2018).  The latter study found that the two-color FUV-optical sequence of NGC~2281 was slightly offset
from that of the Hyades and Coma Ber, suggesting that it was younger than those two comparison clusters.
However, this conclusion was tentative given the uncertainty in reddening.  K13 measure a radius of 24\farcm3 so the UVOT images
only cover the center of this low-density cluster.

The astrometric membership includes one spectroscopic member from MMU and defines a clear sequence in the CMDs
(Figure \ref{f:CMDfig1}, third column) that corresponds almost exactly to the consensus literature values, with slightly higher reddening ($E(B-V)=0.15$). That the reddening is slightly higher than 
the value assumed in Smith (2018) would allow
NGC~2281's two-color sequence to better overlap that of the Hyades and Coma Ber (see their Figure 5, which assumes $E(B-V)=0.06$). This would be consistent with our estimate
of an NGC~2281 age similar to that of the Hyades and Coma Ber.

\subsubsection{M~41 (NGC~2287)}

NGC~2287 (M~41) is a bright nearby second quadrant cluster situated below the Galactic midplane.  Extensive study has shown it to be young (200 Myr)
minimally
reddened and with a high binary fraction (Harris et al. 1993, Dobbie et al. 2012).  The UVOT field only covers the central region of the cluster (K13 measure a radius of 26\farcm4) and the data suffered
from massive saturation from its many bright stars, which prevented both adequate PSF fitting and photometry of some of the brightest cluster
members.  It also created numerous false detections around the image artifacts but these were easily removed by matching
to the GAIA DR2 catalog.

The astrometric selection includes a number of radial velocity members from MMU.
Selecting these stars and examining their aperture photometry shows a clear narrow main sequence which is well-fit by the
M17 isochrones with parameters similar to those derived by previous investigators (Figure \ref{f:CMDfig1}, fourth column).  The saturation limits our analysis, allowing
us to only place an upper limit (500 Myr) on the age of M~41 and making it impossible to confirm the eMSTO report by Cordoni et al. (2018). We detect a number of
confirmed astrometric members along a probable white dwarf sequence and show the deviation in the $(uvw1-uvw2)$ colors seen in several other clusters.

\subsubsection{NGC~2301}

\begin{figure}[h]
\begin{center}
\includegraphics[scale=.5,angle=270]{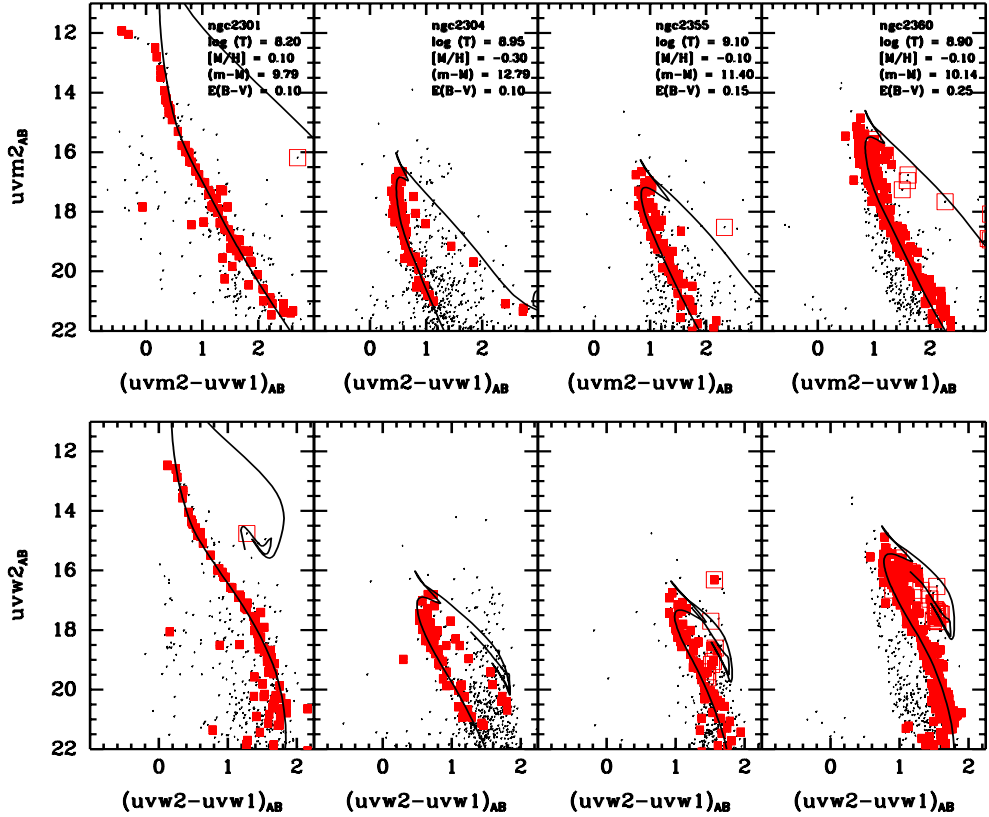}
\end{center}
\caption{Color-magnitude diagrams of the open clusters NGC~2301, NGC~2304, NGC~2355 and NGC~2360 (left to right). The solid lines are PARSEC-COLIBRI isochrones
set to the parameters in the text.  Solid red squares show astrometric members
selected from GAIA DR2 by CG18 while open squares are spectrosopic members from either MMU or sources given in the text.\label{f:CMDfig2}}
\end{figure}

NGC~2301 is a bright third quadrant cluster located within the Galactic midplane.  The catalogs of K16 and Paunzen et al. (2010) describe it as
slightly metal-rich ($[M/H]=+0.06$) and with a small amount of foreground reddening ($E(B-V)=0.06$).  It has been extensively studied for variable stars (see, e.g., Wang et al. 2015).
As with NGC~2287, which it bears a striking
similarity to, there is a plethora of bright
saturated stars prevented PSF-fitting photometry.  The data presented is aperture photometry cleaned by matching to the GAIA DR2 and corrected to the same aperture
as the PSF photometry.

The astrometrically-selected CMD (Figure \ref{f:CMDfig2}, first column) shows clear loci for the cluster stars.  The one member star catalogued by MMU is well within the proper motion
locus and along the RGB, albeit significantly fainter than expected.
The M17 isochrones fit the photometry quite well, providing an upper limit on the age of 400 Myr and favoring a slightly higher foreground reddening ($E(B-V)=0.10$).

\subsubsection{NGC~2304}

NGC~2304 is a third quadrant cluster located above the Galactic midplane.  Multiple photometric studies have been done, producing a broad range
of parameters as shown in Table 2 (Ann et al. 2002, Hasegawa et al. 2008,  Lata et al. 2010, Oralhan et al. 2015). We were unable to identify a comprehensive radial velocity catalog
and the cluster was not surveyed by MMU.

The astrometric selection produces a clean CMD (Figure \ref{f:CMDfig2}, second column).  The M17 isochrones favor a cluster that is intermediate age (900 Myr), moderately 
metal-poor ($[M/H]=-0.3$) and slightly reddened ($E(B-V)=0.10$).  Interestingly, our fit is slightly improved if we use an SMC reddening law (with no blue bump) as opposed to a Milky Way reddening
law, although the difference is very small at such a low level of foreground reddening.

\subsubsection{NGC~2343}
\label{ss:ngc2343}

NGC~2343 is a third quadrant cluster situated near the Galactic midplane.  Pe{\~n}a \& Mart{\'{\i}}nez (2014) showed the cluster to be young (12 Myr) and {\it metal-poor}, with
$[Fe/H]\sim-0.4$ based on $uvby-\beta$ photometry. It would be unusual for such a young cluster to be so metal-poor.  Netopil et al. (2016), by contrast, estimate
a photometric metallicity if $[Fe/H]=-0.03$.  Several catalogues, including WEBDA, list the cluster at $[Fe/H]=-0.3$ but these appear to all trace back to a photometric metallicity
estimate of $[Fe/H]=-0.2$ from Claria (1985).

\begin{figure}[h]
\begin{center}
\includegraphics[scale=.5,angle=270]{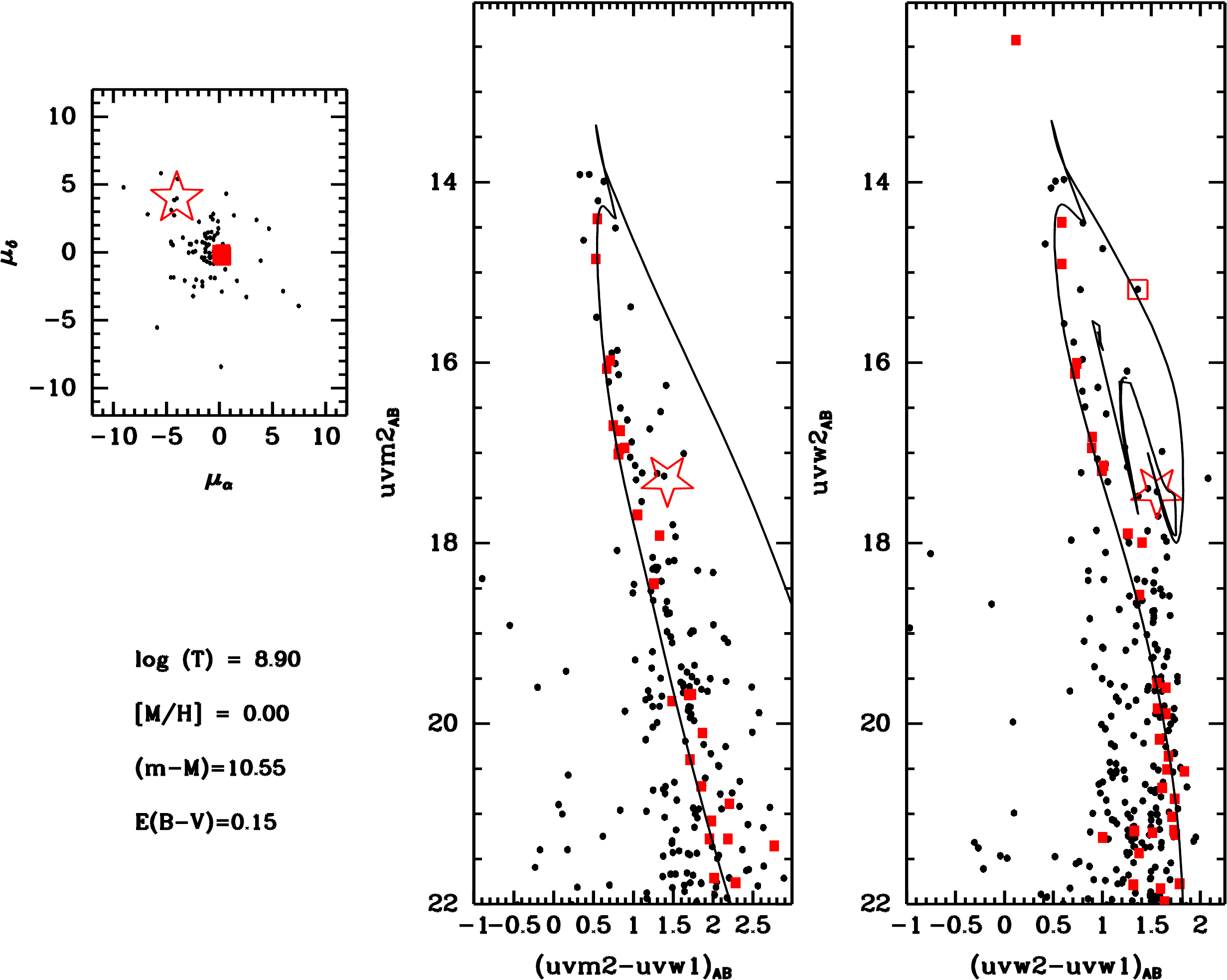}
\end{center}
\caption{Vector-point and color-magnitude diagrams of the cluster NGC~2343 compared to the best-fit isochrones of PARSEC-COLIBRI. The brightest
stars in the cluster are saturated, preventing us from measuring more than an upper limit on the age of the cluster.
The solid red squares show astrometric members
selected from GAIA DR2 while the open red square is a spectroscopic member from MMU. The starred point marks the star used by  Pe{\~n}a \& Mart{\'{\i}}nez (2014) to estimate
the metallicity of the cluster.\label{f:NGC2343}}
\end{figure}

For unknown reasons, the CG18 catalog of NGC~2343 was among a few third-quadrant clusters that had very few stars with high cluster membership probabilities.
We therefore created our own membership probability measures
by fitting two Gaussians to the distribution of proper motions in RA-DEC space. Cluster membership was defined as the ratio of the two Gaussians and members with ratios greater than
9.0 were identified as likely members.  This method was initially applied to all of our program clusters. For most, it produced member catalog almost identical to CG18 and we chose
to utilized the CG18 membership as it used more information and its membership lists were slightly more conservative. NGC~2343 was one of the few where the methods disagreed.

Figure \ref{f:NGC2343} shows the vector-point diagram of the cluster and shows a tight clump of stars toward the edge of the field star distribution.  Using these
as cluster members, we find that the
metallicity estimate of and Pe{\~n}a \& Mart{\'{\i}}nez are likely in error.  The star they identify as 29 and use as their metallicity estimator
is almost certainly not a member of the cluster, being placed well to the edge of the field star distribution and off of the MS.  It is most likely a metal-poor thick disk or halo star.
HD 54387, used as the metallicity estimator by Claria (1985), is saturated in the UVOT images. Its proper motion does place it near the NGC~2343 proper motion, however.

The member stars trace a clear MS.  Given the uncertainty in the metallicity, we adopt solar metallicity.  If the cluster is indeed slightly metal-poor, the result would be a shorter
distance modulus ($m-M=9.64$).
While our photometry is limited in its ability to measure the age of this cluster and we can only give an upper limit (the bright star seen in the $uvw2-uvw1$ CMD is saturated),
we find it unlikely that NGC~2343 is a very young
cluster, given the small number of bright saturated stars in the images and the location of the spectroscopically-confirmed RGB star from MMU. The location of the RGB star,
approximately three magnitudes brighter than the MS, is consistent with an intermediate-age population.  A young population would have a much hotter and brighter RGB sequence.
It is therefore likely that the cluster age is more in the range of 100-500 Myr, as indicated by our photometry and Netopil et al., than the 12 Myr measured by 
Pe{\~n}a \& Mart{\'{\i}}nez.

\subsubsection{NGC~2355}

NGC~2355 is a third quadrant cluster located above the Galactic midplane.  Previous investigations have revealed it to be slightly
metal-poor ($[M/H]=-0.06$) and of intermediate age (900 Myr, Donati et al. 2015) with typical disk abundance ratios (Jacobson et al. 2011).

The CMDs (Figure \ref{f:CMDfig2}, third column) show a fit consistent with previous investigations, but favoring a slightly higher age of 1.1 Gyr.
The MSTO of the cluster proved difficult to fit precisely with any combination of parameters. Notably, we find that there is some broadening in the MSTO.
This could indicate an eMSTO but could also  be the effect of differential reddening (to which the NUV is particularly sensitive at the level of $E(B-V)=0.05$).
The cluster may be worthy of future spectroscopic study to determine the nature of the MSTO broadening.

\subsubsection{NGC~2360}
\label{ss:ngc2360}

NGC~2360 is third quadrant cluster located above the Galactic midplane.
The literature on this cluster contains widely different values for the parameters as shown in Table 2.
Reddy et al. (2012) and Sales-Silva et al. (2014) favor a large distance ($m-M=11.72$) and young age (560 Myr) while Claria et al. (2008) favors a closer distance ($m-M=10.09$)
and older age (1.8 Gyr).
This discrepancy is likely due to the presence of two distinct color-magnitude
sequences in the field, something we have observed in numerous low-latitude third quadrant clusters, as detailed in a number of following clusters.  One of these represents
foreground disk; the other the cluster itself.  WEBDA lists the cluster at the longer distance and younger age.

Previous spectroscopic studies have been mostly focused on red giants, which our data cannot use to constrain the properties of the cluster, due to their lack of UV emission.
However, the astrometric selection (Figure \ref{f:CMDfig2}, last column) clearly aligns with the brighter/redder main sequence. This sequence is favored independently by the proper motions, parallaxes and spatial
distribution.  Fitting the M17 isochrones to this sequence produces a fit consistent with the Claria et al. result and
inconsistent with the Reddy/Sales-Silva result current listed in WEBDA. We conclude that the closer/older solution is likely the correct one.

\begin{figure}[h]
\begin{center}
\includegraphics[scale=.5,angle=0]{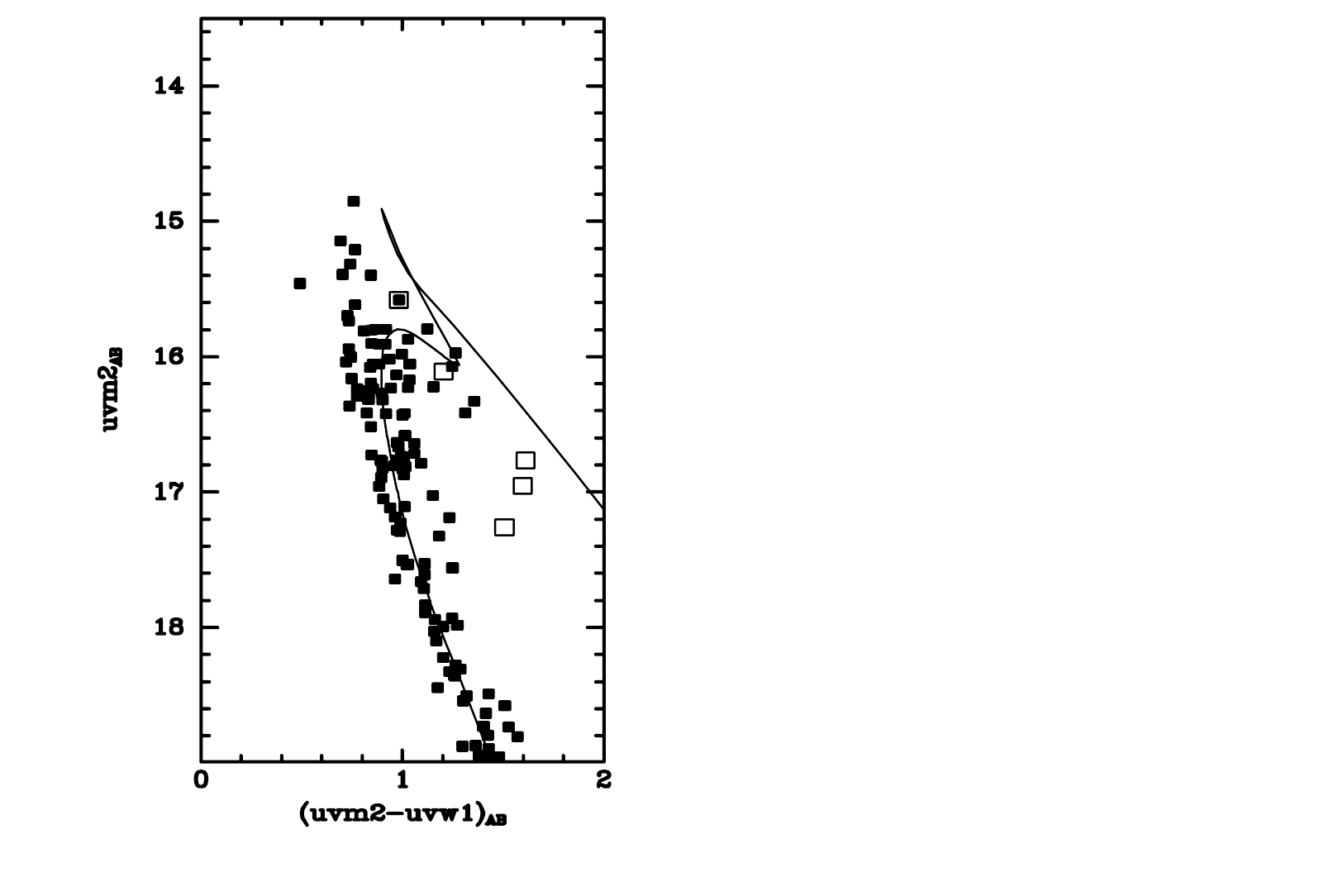}
\end{center}
\caption{A close-up of the MSTO region of NGC~2360.  The panel shows GAIA- and spectroscopically-selected members. The MSTO region is very broad due to the combination
of eMSTO, convective hook and the beginning of the RGB.\label{f:NGC2360msto}}
\end{figure}

NGC~2360 is one of the few clusters in our sample that is both bright enough and close enough to explore the fine structure of the MSTO.  Milone et al. (2018) identified it as a cluster
with an eMSTO. This region of the CMD is tricky because it overlaps both the convective hook in the MSTO and the beginning of the RGB.  However, the $uvm2-uvw1$ diagram
shows that the MSTO is broader and more structured than one would expect from a simple stellar population (the fit isochrone splits the difference between the blue and red edges of the MSTO).
Figure \ref{f:NGC2360msto} shows a close-up of this turnoff region and it appears to confirm the Milone et al. (2018) discovery, hinting at a bifurcation in the MSTO, likely as a result of variation
in stellar rotation.

\subsubsection{Berkeley 37}

\begin{figure}[h]
\begin{center}
\includegraphics[scale=.5,angle=270]{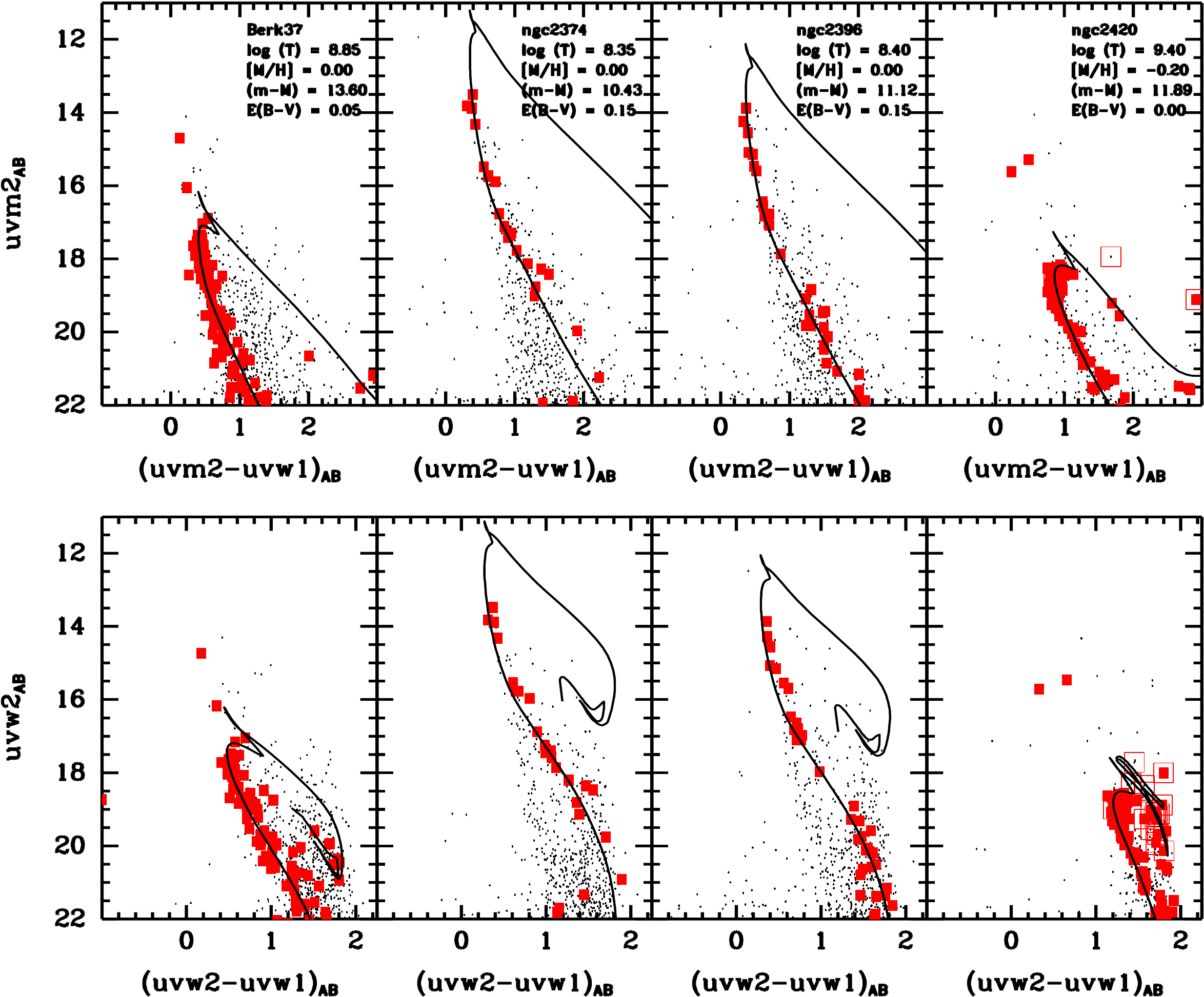}
\end{center}
\caption{Color-magnitude diagrams of the open clusters Berkeley~37, NGC~2374, NGC~2396 and NGC~2420 (left to right). The solid lines are PARSEC-COLIBRI isochrones
set to the parameters in the text.  Solid red squares show astrometric members
selected from GAIA DR2 by CG18 while open squares are spectrosopic members from either MMU or sources given in the text.\label{f:CMDfig3}}
\end{figure}

Berkeley 37 is a faint third quadrant cluster situated above the plane.  The only photometric study to date is that Oralhan et al. (2015) who
showed to be distant ($m-M=13.60$) with light extinction ($E(B-V)=0.05$) and an intermediate age (600 Myr).  K16 derive a slightly shorter distance with more reddening.

The CG18 selection did not pick out many cluster members, likely due to the similarity of the cluster kinematics to that of the disk. We applied our own selection criteria using the double-Gaussian method
described for NGC~2343 (\S \ref{ss:ngc2343}).  The VPD shows a tight clump of stars within the dense field star distribution.  This clump corresponds to a faint intermediate-age main sequence
in the CMD (Figure \ref{f:CMDfig3}, first column).  We measure cluster parameters similar to that of Oralhan.  Our membership survey includes two stars that
could be potential blue stragglers.  One of these in the core of the cluster while the other is in the outskirts.  This indicates that the former
is likely a genuine blue straggler while the latter may be a field star contaminant for a cluster in a very dense region of the sky.

\subsubsection{NGC~2374}

NGC~2374 is third quadrant cluster situated near the Galactic midplane.  It has been poorly studied with no spectroscopic metallicity available. The only recent
photometric study is Carraro et al. (2015), who found it be young (250 Myr) and argued that its position near the Galactic warp makes it consistent with either a thick disk cluster
or an extended thin disk cluster.

MMU only identified one member star -- a red giant that is faint and well-removed from the main sequence.  CG18 only identify
a handful of astrometric members.  The VPD, however, shows
a clear clump of stars well-removed from the disk sequence that defines narrow photometric sequences in the CMDs (Figure \ref{f:CMDfig3}, second column).  We derive parameters similar to
that of Carraro et al. (2015) albeit at slightly higher reddening ($E(B-V)=0.15$). Like many third quadrant clusters, we also see the ``second sequence"
of the disk.

\subsubsection{NGC~2396}

NGC~2396 is a third quadrant positioned just above the Galactic midplane. It has never been the focus of an individual study and the only parameters
in the literature come from the comprehensive survey of K16, who showed the cluster to be nearby ($m-M=8.68)$, of intermediate age (300 Myr) and minimally reddened
($E(B-V)=0.05$).

The CG18 astrometric selection produces a clear sequence in CMDs (Figure \ref{f:CMDfig3}, third column).  However, the main sequence is over 2.5 mag fainter
than that predicted by the K16 parameters.  The best fit shows a similar age (250 Myr), slightly more reddening ($E(B-V)=0.15$) and a dramatically larger distance
($m-M=11.12$).
Interestingly, the faint end of the $uvw2-uvw1$ sequence does not show the deviation between data and isochrone that other bright clusters show even though
the underlying parameters of the stellar population are similar.

\subsubsection{NGC~2420}

NGC~2420 is a third quadrant cluster positioned 19 degrees above the Galactic midplane.  Extensive investigation (Von Hippel \& Gilmore 2000, Anthony-Twarog et al. 2006, Souto et al. 2016) 
has revealed it to be old (2 Gyr) and slightly metal-poor ($[Fe/H]\sim-0.14$).  It is also one of the few clusters
to be extensively studied in the UV by GALEX (De Martino et al. 2008) and has extensive radial velocity data in MMU.

Fitting the M17 isochrones to the astrometrically-selected stars using literature values (Figure \ref{f:CMDfig3}, last column) produces a reasonable fit but with a slight discrepancy: the $uvm2-uvw1$ model colors
are a little too red and the $uvw2-uvw1$ colors are a little too blue.  A detailed examination of the photometry, including a direct comparison
to the standard Swift/UVOT pipeline photometry produced by UVOTSOURCE shows no error in the photometry.  NGC~2420 give no indication of any chemical abundance anomalies, showing typical metal-rich
disk abundances with solar-type $\alpha$-abundances (Suoto et al. 2016).  Lowering the metallicity to [Fe/H]=-0.3 and decreasing the cluster distance produces
very good isochrones fits.  However, the abundance of NGC~2420 is very-well stablished through multiple spectroscopic studies.
Previous research has indicated some variation in the UV
reddening law within external galaxies and within the Galaxy, particularly the $R_V$ value and bump strength (Siegel et al. 2014, Hagen et al. 2017) but the foreground
reddening in NGC~2420 is low and modifying the reddening law produces only minor changes. We are unable, at this point, to explain the slight discrepancy in
NGC~2420 compared to other clusters that have better fits.

We do note, however, that the MSTO does show a bit of broadening, which would be consistent with an eMSTO.  NGC~2420's 2 Gyr age is toward the lower end of the range where eMSTO
is expected to manifest before magnetic breaking evens out the stellar rotation rates (Georgy et al. 2019).  It would be worth further
investigation to determine if the MSTO stars shows differences in rotation rates and potentially set the age limits of the eMSTO phenomenon on a more empirical footing.

\subsubsection{NGC~2422}

\begin{figure}[h]
\begin{center}
\includegraphics[scale=.5,angle=270]{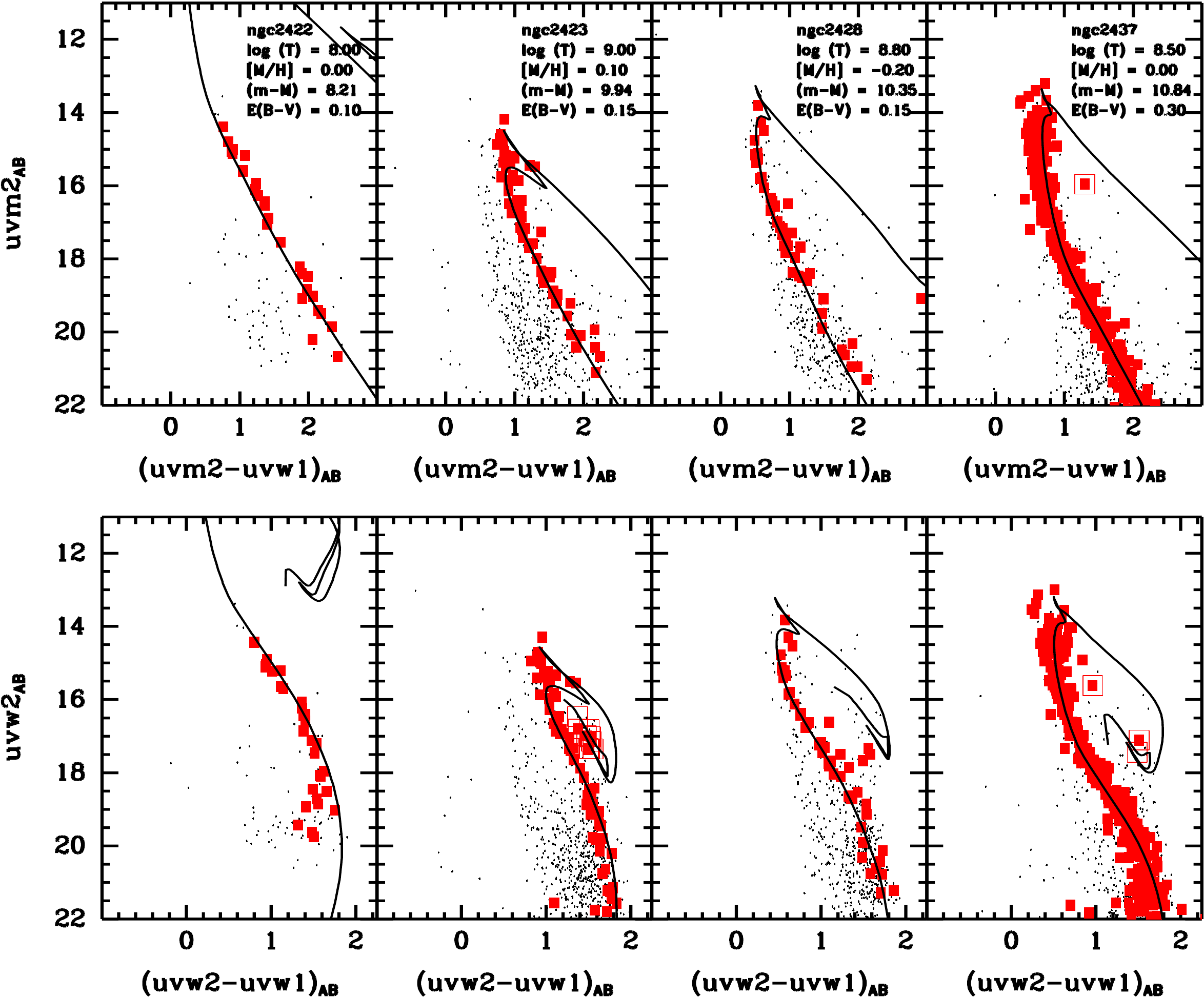}
\end{center}
\caption{Color-magnitude diagrams of the open clusters NGC~2422, NGC~2423, NGC~2428 and NGC~2437 (left to right). The solid lines are PARSEC-COLIBRI isochrones
set to the parameters in the text.  Solid red squares show astrometric members
selected from GAIA DR2 by CG18 while open squares are spectrosopic members from either MMU or sources given in the text.\label{f:CMDfig4}}
\end{figure}

NGC~2422 is a bright nearby third quadrant cluster positioned near the Galactic midplane.  Studies by Rojo Arellano et al. (1997)
and Prisinzano et al. (2003)
have shown it to be very young (80 Myr) with moderate foreground reddening ($E(B-V)=0.09$). Conrad et al. (2014) measure it as slightly metal poor ($[Fe/H]=-0.03$)
but within uncertainties of solar metallicity.
The cluster proved difficult to study due to the large number of saturated stars
and the concomitant number of false detections in the diffraction spikes. The sample studied was only of stars the could be matched the GAIA DR2. The cluster covers a large
area of sky -- 37\farcm8 in radius according to K13 -- and thus we have only imaged the core.

The CMDs (Figure \ref{f:CMDfig4}, first column) shows a narrow young MS consistent with literature expectations.  The brightest stars in the
cluster are badly saturated, meaning we can only place an upper limit on the age.  However, the reddening and distance are well-constrained. As with other clusters, the
faint end of the $uvm2$-$uvw1$ sequence shows a deviation from the isochrone, bending more sharply toward the blue than predicted.

\subsubsection{NGC~2423}

NGC~2423 is a third quadrant cluster positioned above the Galactic midplane. Prior study has revealed it be a intermediate age (1 Gyr) and metal-rich ($[Fe/H]=+0.10$) cluster
(Claria et al 2008, Santos et al. 2009, 2012, Paunzen et al. 2010, Conrad et al. 2014).

Like NGC~2360 (\ref{ss:ngc2360}, the CMD shows a double sequence of the old thin disk and cluster (Figure \ref{f:CMDfig4}, second column).  The astrometric selection picks the brighter
sequence and contains the radial velocity members identified by MMU. 
The M17 isochrones produces a fit with parameters consistent with the literature.

\subsubsection{NGC~2428}

NGC~2428 is a third quadrant cluster positioned 16 degrees below the Galactic midplane.  Conrad et al. (2014) show the cluster to be
slightly metal-poor ($[Fe/H=-0.15$) with modest reddening ($E(B-V)=0.05$).  The global survey of K16 lists a 
shorter distance with heavier foreground reddening.

As with NGC~2360 and NGC~2423, the CMD (Figure \ref{f:CMDfig4}, third column) shows a double sequence and the WEBDA parameters, taken from Conrad et al, appear
to trace the bluer sequence.  The member stars favor the redder sequence and the M17 isochrones give a fit more consistent with that of K16 than
Conrad et al., indicating that this is the cluster sequence.

\subsubsection{NGC~2437}

NGC~2437 (M~46) is a bright cluster in the third quadrant.  Davidge (2013) shows it to be near solar metallicity, intermediate age (200 Myr)
and moderately reddened.  The Conrad et al. (2014) survey identifies is as very metal-poor, with [M/H] of -0.75, albeit with a very high uncertainty. It would
be unusual, however, for such a young cluster to be so metal-poor.  K13 measure a radius of 34\farcm2 indicating that UVOT only images the core of the cluster.

The astrometric selection of CG18 includes the radial velocity members identified by MMU and traces a long narrow
MS in the CMDs (Figure \ref{f:CMDfig4}, fourth column).  As noted above, the UV CMDs are not very adept at measuring metallicity and NGC~2437 is a case in point.
The properties of the cluster are well-constrained at both low and solar-metallicity.  At solar metallicity, we derive fits close to the parameters of Davidge (2013) and K16,
with a slightly higher reddening ($E(B-V)=0.30$) and shorter distance ($m-M=10.84$).  If the cluster were indeed as a metal-poor as indicated in Conrad et al. (2014), an isochrone
with significantly shorter distances modulus ($m-M=9.96$) and dramatically higher reddening ($E (B-V) = 0.35$) would also fit the data quite well.  Given the high uncertainy in the
latter measure and the lower likelihood that such a young cluster would be so metal-poor, we list NGC~2437's parameters in Table 2 as derived for the higher metallicity.

\subsubsection{NGC~2447}

\begin{figure}[h]
\begin{center}
\includegraphics[scale=.5,angle=270]{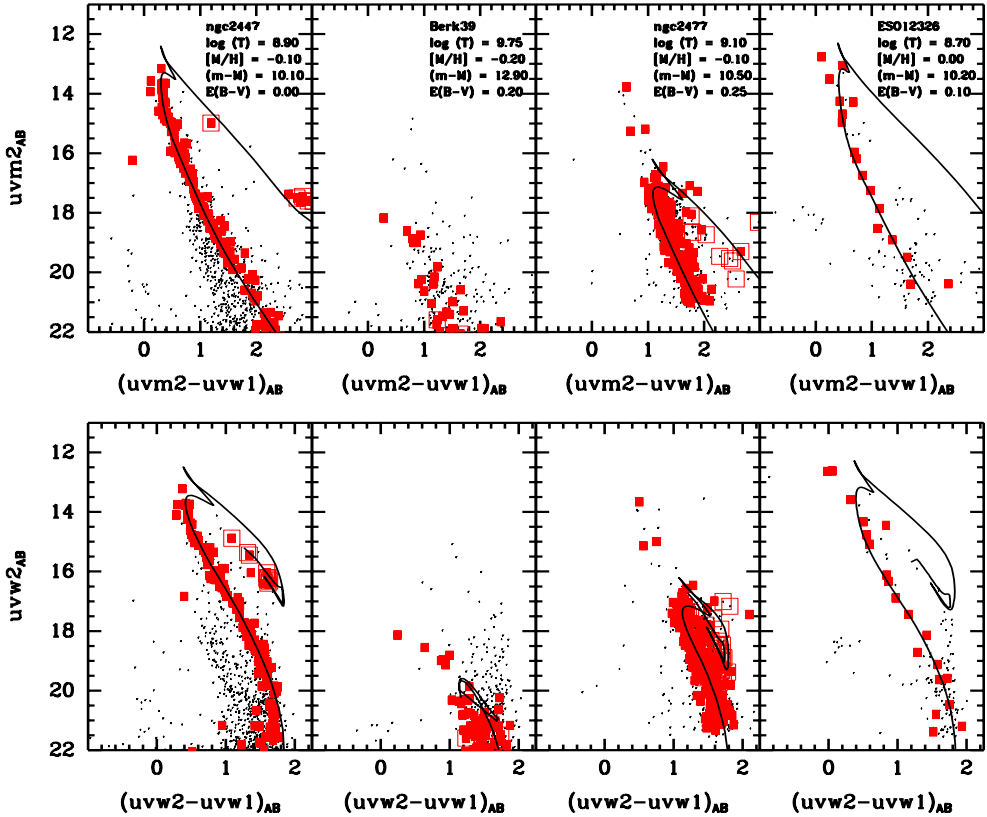}
\end{center}
\caption{Color-magnitude diagrams of the open clusters NGC~2447, Berkely~39, NGC~2477 and ESO~123-26 (left to right). The solid lines are PARSEC-COLIBRI isochrones
set to the parameters in the text.  Solid red squares show astrometric members
selected from GAIA DR2 by CG18 while open squares are spectrosopic members from either MMU or sources given in the text.\label{f:CMDfig5}}
\end{figure}

NGC~2447 is a third quadrant cluster situated within the Galactic midplane.  Studies by Claria et al. (2005), Santos et al. (2009, 2012), Conrad et al. (2014) and Reddy et al. (2015)
describe as having slightly subsolar metallicity ($[Fe/H]=-0.10$) and intermediate age (400 Myr). The K13 radius of 30\farcm0 indicates we observe only the core of the cluster.

Like many other third quadrant clusters, NGC~2447 shows two sequences in the CMD (Figure \ref{f:CMDfig5}, first column).  The astrometric selection, however, favors the
bright sequence of young stars.  The M17 isochrones fit this sequence with parameters very similar to the literature values.

\subsubsection{Berkely 39}

Berkeley~39 is a third quadrant cluster situated 10 degrees above the Galactic midplane.  Previous studies (Kassis et al. 1997; Carraro et al. 1994, 1999; Frinchaboy et al. 2006)
have shown the cluster to be massive, old (5.5 Gyr) and moderately metal-poor ($[Fe/H]=-0.20$)
The extensive spectroscopic survey
of Bragaglia et al. (2012) identifies 30 radial-velocity cluster members with very small star-to-star variations in abundance, consistent with a simple stellar population.  Unfortunately,
only a handful of their program stars have counterparts in the UVOT data due to their cool temperatures and subsequent low UV emission.

Because the cluster is old, reddened and distant, we only detect the very top of the MS in the $uvw2-uvw1$ CMD (Figure \ref{f:CMDfig5}, second column) and
do not detect it all in the $uvm2-uvw1$ CMD (most likely due to the lack of red leak in $uvm2$).
We are unable to constrain its properties on so tenuous a basis.  However, overlaying isochrones from the consensus literature shows a rough agreement.  Moreover,
we detect a number of likely cluster members that would be along a potential blue straggler sequence.

\subsubsection{NGC~2477}
\label{ss:ngc2477}

NGC~2477 is a third quadrant cluster located below the Galactic midplane.  Extensive previous investigation has shown it to be of intermediate age (1 Gyr) with moderate
($E(B-V)=0.25$) reddening (Jeffery et al. 2016)  Metallicity
estimates range from [Fe/H]=+0.07 (Bragaglia et al. 2007) to -0.19 (Conrad et al. 2014). We have taken the metallicity at an intermediate value of -0.10. The cluster
is large, with a K13 radius of 27\farcm0, indicating we only observe the core.

The astrometric and spectroscopic selection identify a clear MS in the CMDs (Figure \ref{f:CMDfig5}, third column).  The M17 isochrones produce a fit
consistent with the previous literature. However, we note that the MS is fairly broad and that this breadth occurs along the entire observed MS. It is likely
this represents differential reddening as NGC~2477 is one of the few clusters we have that has moderate foreground reddening ($E(B-V)=0.25$) and even
small variations can produce noticeable breadth in the CMD. We note at least three stars that are likely blue stragglers.

As noted above, we mostly ignore the RGB stars in the clusters due to their faintness and the dominance of the red leak in their photometric measures.
NGC~2477, however, is one of the that has a prominent RGB. It is also well-surveyed spectroscopically.  One of the most noticeable aspects
of its CMD is that the RGB stars tend to land significantly blueward of the isochrone prediction.  This tendency of the isochrone to miss the few RGB stars
can be seen in other clusters but this the clearest example.
This suggests that the isochrones are not reproducing the NUV properties of the RGB accurately.  The reasons for this could be many but the two most likely are that the atmospheric models do a poor
job of predicting the intrinsically low UV flux for such cool objects (likely due to opacity issues at low temperatures) or that our accounting for the red leak is incorrect.
Further NUV investigation of bright field RGB stars would provide better constraint on both of these inputs of the theoretical isochrones.

\subsubsection{ESO~123-26}

ESO~123-26 is a fourth quadrant cluster positioned well-below the Galactic midplane. It has not been the subject of individual study, with the only parameter
estimates coming from global surveys such as K13 and K16.

This cluster is not included in the CG18 compilation and so we applied the double Gaussian selection method used for NGC~2343 (\S \ref{ss:ngc2343}).
The VPD shows a loose clumping of stars toward the outside of the field star distribution which corresponds to a clear color-magnitude sequence (Figure \ref{f:CMDfig5}, last column).
The best-fit
isochrone is close to the K16 parameters but with a longer distance ($m-M=10.20$) and a slightly older age (500 Myr). We caution however, that the proximity of this cluster
may mean that its brightest members are saturated, so this age should be regarded as an upper limit.

\subsubsection{NGC~2479}

\begin{figure}[h]
\begin{center}
\includegraphics[scale=.5,angle=270]{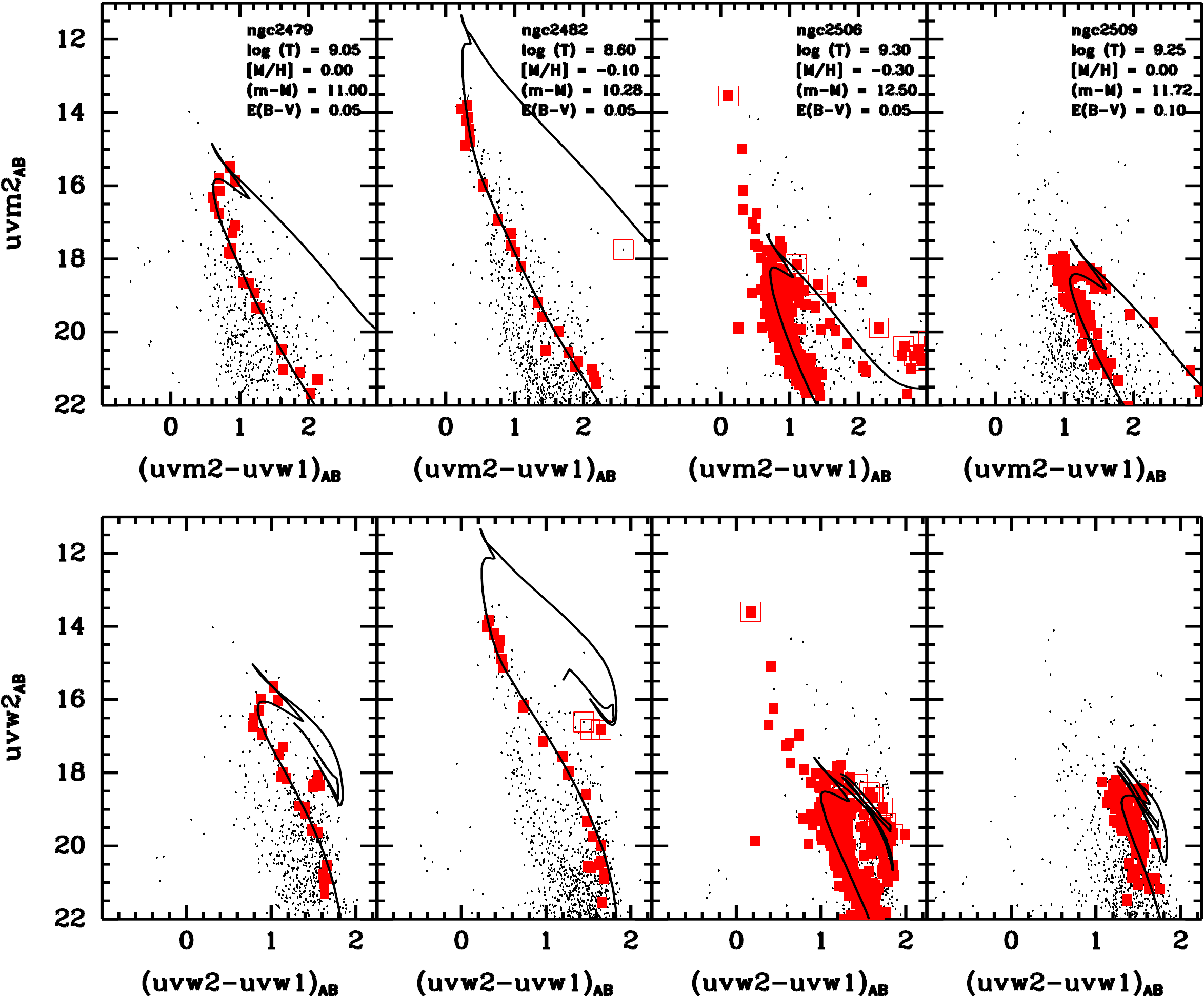}
\end{center}
\caption{Color-magnitude diagrams of the open clusters NGC~2479, NGC~2482, NGC~2506 and NGC~2509 (left to right). The solid lines are PARSEC-COLIBRI isochrones
set to the parameters in the text.  Solid red squares show astrometric members
selected from GAIA DR2 by CG18 while open squares are spectrosopic members from either MMU or sources given in the text.\label{f:CMDfig6}}
\end{figure}

NGC~2479 is above the Galactic midplane in the fourth quadrant.  A recent photometric study by Piatti et al. (2010) revealed the cluster to be old (1 Gyr) and relatively distant 
($m-M=11.00$) with minimal foreground reddening.

NGC~2479 is also not included in the CG18 study and we applied our own astrometric selection (see \S \ref{ss:ngc2343}).
The VPD reveals a tight clump of stars toward the edge of the field star distribution.  Stars selected from this clump trace a narrow
sequence in the CMD (Figure \ref{f:CMDfig6}, first column).  The parameters of Piatti et al. (2010) produce isochrones that overlay this sequence
almost exactly with a slightly older age of 1.1 Gyr.

\subsubsection{NGC~2482}

NGC~2482 is a fourth quadrant cluster positioned just above the Galactic disk.  Recent studies
have shown it to be an intermediate age (400 Myr) near-solar metallicity cluster (Reddy et al. 2013, Krisciunas et al. 2015,
Conrad et al. 2014).
The astrometric selection shows a clear sequence in the color-magnitude diagram (Figure \ref{f:CMDfig6}, second column).  The M17 isochrones produce
a fit similar to previous studies but with lower reddening ($E(B-V)=0.05$).

\subsubsection{NGC~2506}

NGC~2506 is a third quadrant cluster located above the Galactic midplane.  Previous studies have shown it to be an old (2 Gyr) metal-poor ($[Fe/H=-0.3$) cluster
(Reddy et al. 2012b, Lee et al. 2012. Anthony-Twarog et al. 2016). 
The cluster has also been the subject of extensive spectroscopic surveys which have confirmed its low metallicity (MMU, Anthony-Twarog et al. 2018).

The CMD shows a dominant locus of cluster stars, which contains the radial velocity members of MMU and astrometrically-selected members (Figure \ref{f:CMDfig6}, third column).  The M17 isochrones
produce a fit consistent with prior literature given the assumed metallicity of [Fe/H]=-0.3.  NGC~2506 also shows a very prominent blue straggler sequence, the brightest of which is both a proper motion
and radial velocity member. There is quite a bit of breadth to the MS, which is unexpected given the low reddening.  The broadening occurs along the entire length of the MS rather than just
the MSTO, making the nature of it unclear.

\subsubsection{NGC~2509}

NGC~2509 is an older cluster positioned above the Galactic disk.
The literature shows a number of very different solutions to NGC~2509's properties.  Sujatha \& Babu (2003) argue for a very old age of 8 Gyr while Tadross (2055) and Carraro \& Costa (2007) 
find a more intermediate age of 1.2-1.6 Gyr, but with a significant difference in distance modulus between them ($m-M=11.50$ and 12.50, respectively).  No spectroscopic
metallicity is available and those studies have assumed solar metallicity.

The CG18-selected CMD (Figure \ref{f:CMDfig6}, last column) shows a faint MS.  Assuming solar metallicity, the best fit to the photometry is
consistent with the Tadross et al. (2005) results, showing an older age (1.7 Gyr) and
a shorter distance ($E(B-V)$=11.72).

\subsubsection{NGC~2527}

\begin{figure}[h]
\begin{center}
\includegraphics[scale=.5,angle=270]{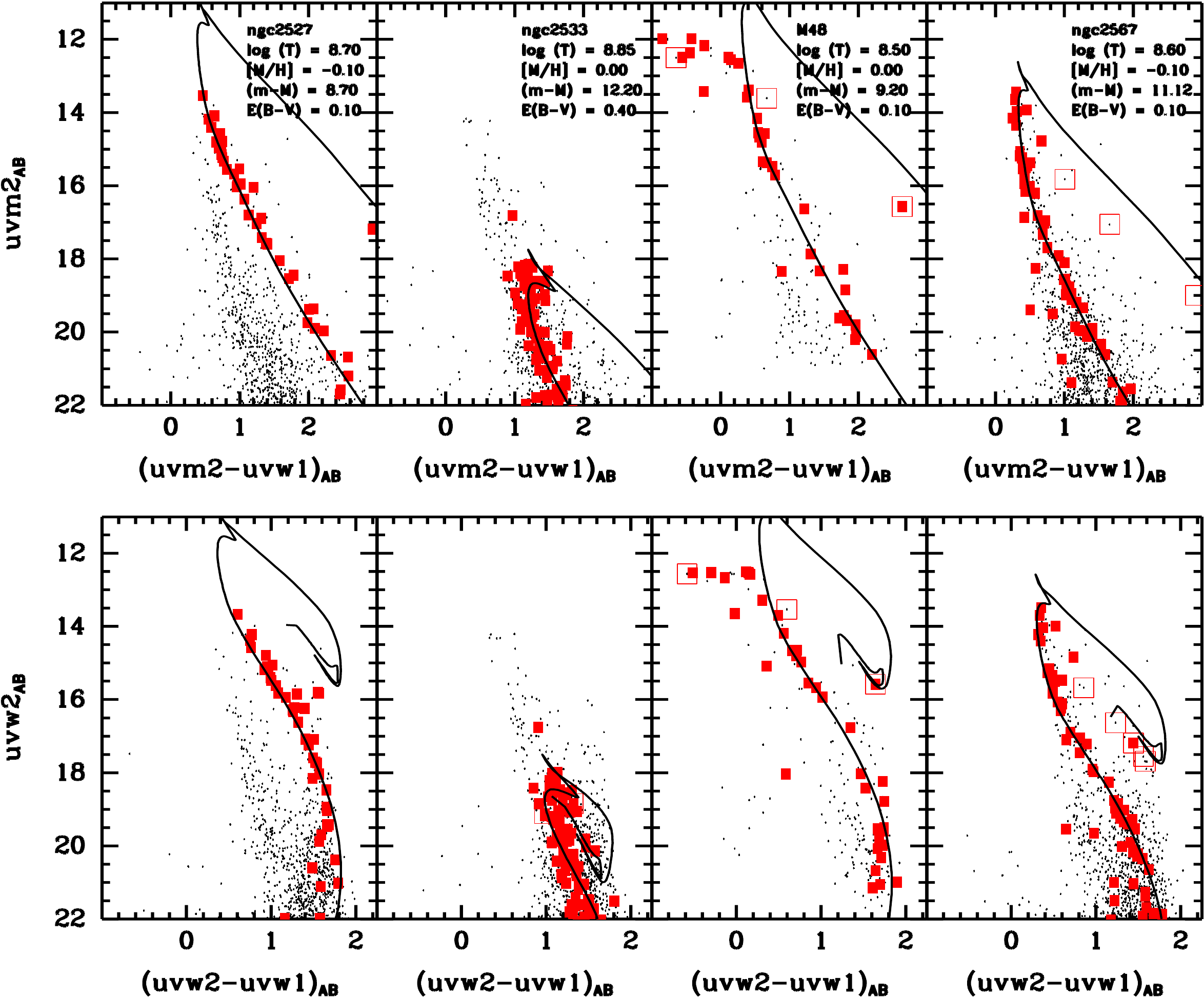}
\end{center}
\caption{Color-magnitude diagrams of the open clusters NGC~2527, NGC~2533, M~48 and NGC~2567 (left to right). The solid lines are PARSEC-COLIBRI isochrones
set to the parameters in the text.  Solid red squares show astrometric members
selected from GAIA DR2 by CG18 while open squares are spectrosopic members from either MMU or sources given in the text.\label{f:CMDfig7}}
\end{figure}

NGC~2527 is a third quadrant cluster near the Galactic midplane.  Reddy et al. (2013) found it be of intermediate age (450 Myr)
slightly metal-poor ($[Fe/H=-0.09$) and slightly reddened ($E(B-V)=0.04$), based on spectroscopy and photometry. Conrad et al. (2014) find it to be slightly more metal-rich
($[Fe/H]=+0.06$) but based only on two stars.

The CMD (Figure \ref{f:CMDfig7}, first column) shows a clean sequence, well-separated from the disk.  We find parameters
similar to those of Reddy et al..  The age of the cluster is a bit uncertain as the brightest MSTO stars are saturated. However, the curvature of the upper parts
of the CMD are inconsistent with an age older than that found in Reddy et al.  Note again that the $uvw2-uvw1$ colors at the faint end of the MS are bluer than predicted by the isochrones.

\subsubsection{NGC~2533}

NGC~2533 is a third-quadrant cluster positioned above the Galactic midplane.  The cluster has never been targeted for individual photometric study.
However, basic parameters come from the blue straggler survey of Ahumada \& Lapasset (2007) and the global survey of K16.

The CG18 selection showed very few members stars so we applied our own proper-motion analysis, as detailed in \S \ref{ss:ngc2343}.
The CMD (Figure \ref{f:CMDfig7}, second column) shows a rough faint main sequence and turnoff
that are similar to the expectations form the K16 parameters. We adjusted those parameters slightly to a shorter distance ($m-M=12.20$) and older age (700 Myr), although we emphasize
that the analysis of this cluster is complicated by the somewhat high ($E(B-V)=0.40$) reddening and red leak causing the RGB to cross over the MSTO in the $uvw2-uvw1$ CMD.

\subsubsection{M~48 (NGC~2548)}

M~48 is one of the most well-studied open clusters in the Galaxy. While the specific parameters of the various studies vary, there is a general consensus that is close, of intermediate
age (400 Myr) and with minimal reddening (Rider et al. 2004; Balaguer-N{\'u}{\~n}ez et al.2005; Wu et al. 2005; Sharma et al., 2006).  Most studies assume near solar metallicity although 
Balaguer-N{\'u}{\~n}ez et al. measure a photometric metallicity of [Fe/H]-0.24. The massive size of the cluster (K13 measure a radius of 43\farcm2) indicates that we have only
surveyed the central regions.

The field contains many saturated stars and so only stars with GAIA DR2 counterparts 
are shown in the CMD (Figure \ref{f:CMDfig7}).   Assuming solar metallicity, we derive similar parameters to the consensus literature, with slightly elevated reddening ($E(B-V)=0.10$).
We note that
the brightest stars in M~48 are saturated in the UVOT data and so we can only give an upper limit.  However, the curvature of the upper MS is consistent with an age of $\sim$ 300 Myr.

\subsubsection{NGC~2567}

NGC~2567 is a third-quadrant cluster positioned slightly above the Galactic midplane.  It 
has not been the subject of individual study. The K16 study provides basic parameters, identifying the clusters as intermediate age (400 Myr) with modest foreground reddening 
($E(B-V)=0.13$) while Conrad et al. (2014) identify it as slightly metal-poor ($[Fe/H]=-0.08$).

The CG18 selection showed very few members stars.  However, our analysis, using the methods of \S \ref{ss:ngc2343}, identifies 
a tight clump of stars in the VPD at the center of the field star distribution.
Lowering the probability threshold to 50\% selects stars that form a clear
MS and MSTO within the CMD (Figure \ref{f:CMDfig7}, last column).  We find that the literature parameters exactly reproduce the color-magnitude sequence.

\subsubsection{NGC~2571}

\begin{figure}[h]
\begin{center}
\includegraphics[scale=.5,angle=270]{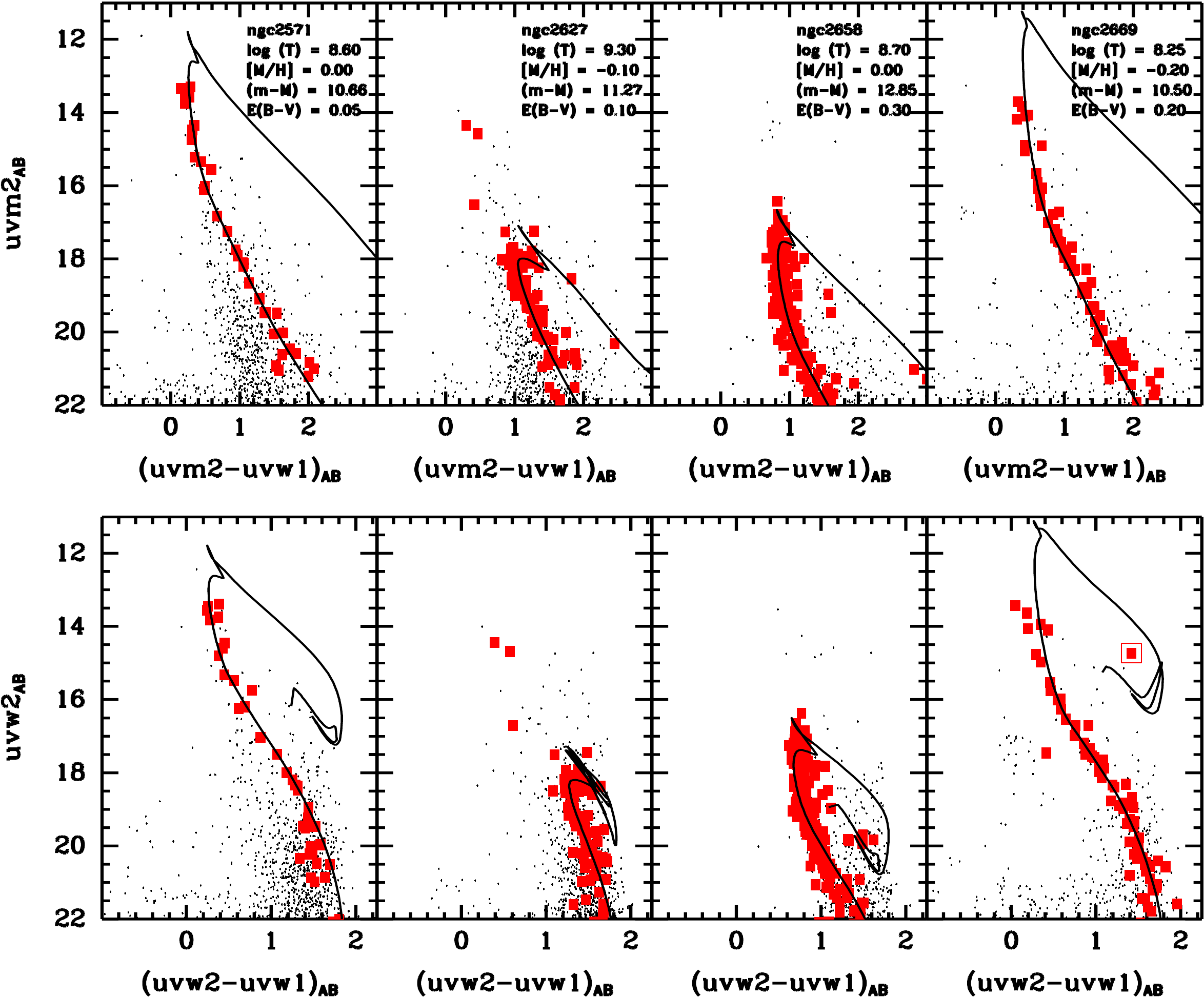}
\end{center}
\caption{Color-magnitude diagrams of the open clusters NGC~2571, NGC~2627, NGC~2658 and NGC~2669 (left to right). The solid lines are PARSEC-COLIBRI isochrones
set to the parameters in the text.  Solid red squares show astrometric members
selected from GAIA DR2 by CG18 while open squares are spectrosopic members from either MMU or sources given in the text.\label{f:CMDfig8}}
\end{figure}

NGC~2571 is a third quadrant cluster positioned above the midplane.  It has
been analyzed by Giorgi et al. (2002) and {\"O}zeren et al. (2014).  The latter, based on 2MASS photometry, shows moderate reddening ($E(B-V)=0.36$) while the former shows a small amount
of foreground reddening ($E(B-V)=0.10$.  Both agree that the cluster is quite young ($\sim50$ Myr).

The CMD (Figure \ref{f:CMDfig8}, first column) shows a clean MS well-separated from the disk.  We find, however,
that the parameters derived by previous investigators are somewhat inconsistent with the NUV photometry.  The high reddening value derived by {\"O}zeren et al. (2014) would be inconsistent
with the color of the upper MS for any age, predicting a $uvm2-uvw1$ color several tenths of a magnitude redward of the photometry.  The very young ages derived by the previous studies are also
inconsistent with the curvature of the upper MS, the brightness of the main sequence and the color of the main sequence. While we have multiple saturated stars in the field and our age
thus represents an upper limit, we note that Kilambia et al. (1978) derive an age of 175 Myr based on photographic photometry, a result that is far more
consistent with the NUV photometry we present here.  We therefore find it unlikely that this cluster is very young.

\subsubsection{NGC~2627}

NGC~2627 is a third quadrant cluster situated just above the Galactic midplane. Prior photometric studies have derived an old age of 1.6 Gyr
(Piatti et al. 2003; Ahumada 2005). There is no spectroscopic
metallicity but Piatti et al. estimate that is slightly subsolar ([Fe/H]$\sim -0.1$) based on Washington indices.

MMU surveyed the cluster but none of their four RGB stars are detected in the Swift data.  The astrometric selection, however, produces
a clear CMD with a well-defined MS and MSTO (Figure \ref{f:CMDfig8}, second column).  The best-fit isochrone roughly corresponds to the parameters from Ahumada.  Note that because of the age,
distance, and reddening
of this cluster, the RGB/AGB region overlaps the MSTO in the $uvw2-uvw1$ region.  The actual MSTO is the faintest of these loops. Several astrometric members are in the blue straggler region
of the CMD, indicating a likely association.

\subsubsection{NGC~2658}

NGC~2658 is a third quadrant cluster located six degrees above the Galactic midplane.
The cluster has not been the target of spectroscopic study.  Paunzen et al. (2010) indicated that it was metal-poor but the updated survey of Netopil (2016) indicate
solar metallicity.  Photometric studies (Ramsay \& Pollaco 1992, Ahumada 2003) describe it as intermediate age (300 Myr), solar metallicity and
with moderate reddening ($E(B-V)=0.40)$, making it one of the few significantly reddened clusters in our sample.

The CMDs show a prominent MS and MSTO (Figure \ref{f:CMDfig8}, third column).  Fitting the M17 isochrones
to this sequence results in parameters similar to the literature, with a slightly lower reddening value of $E(B-V)=0.30$.
The MS does show a significant breadth and some extension beyond the MSTO.  This could indicate
that it has an eMSTO.

\subsubsection{NGC~2669}

NGC~2669 is a poorly-studied third quadrant cluster located six degrees below the Galactic midplane. The last focused study of the cluster
was Vogt \& Moffat (1973) but it has been included in the global surveys of Paunzen et al. (2010), K13 and K16.  These studies hint at a young (100 Myr) slightly metal-poor ($[Fe/H]\sim-0.20$),
moderately reddened ($E(B-V)=0.18$) cluster.

The astrometric selection includes one MMU radial velocity member and traces the bluer sequence in the CMDs (Figure \ref{f:CMDfig8}, last column).
Assuming the Paunzen et al. metallicity, we find parameters similar to those in the literature.  We note, however, that the brightest stars are saturated and so the age given
is an upper limit.

\subsubsection{NGC~2818}

\begin{figure}[h]
\begin{center}
\includegraphics[scale=.5,angle=270]{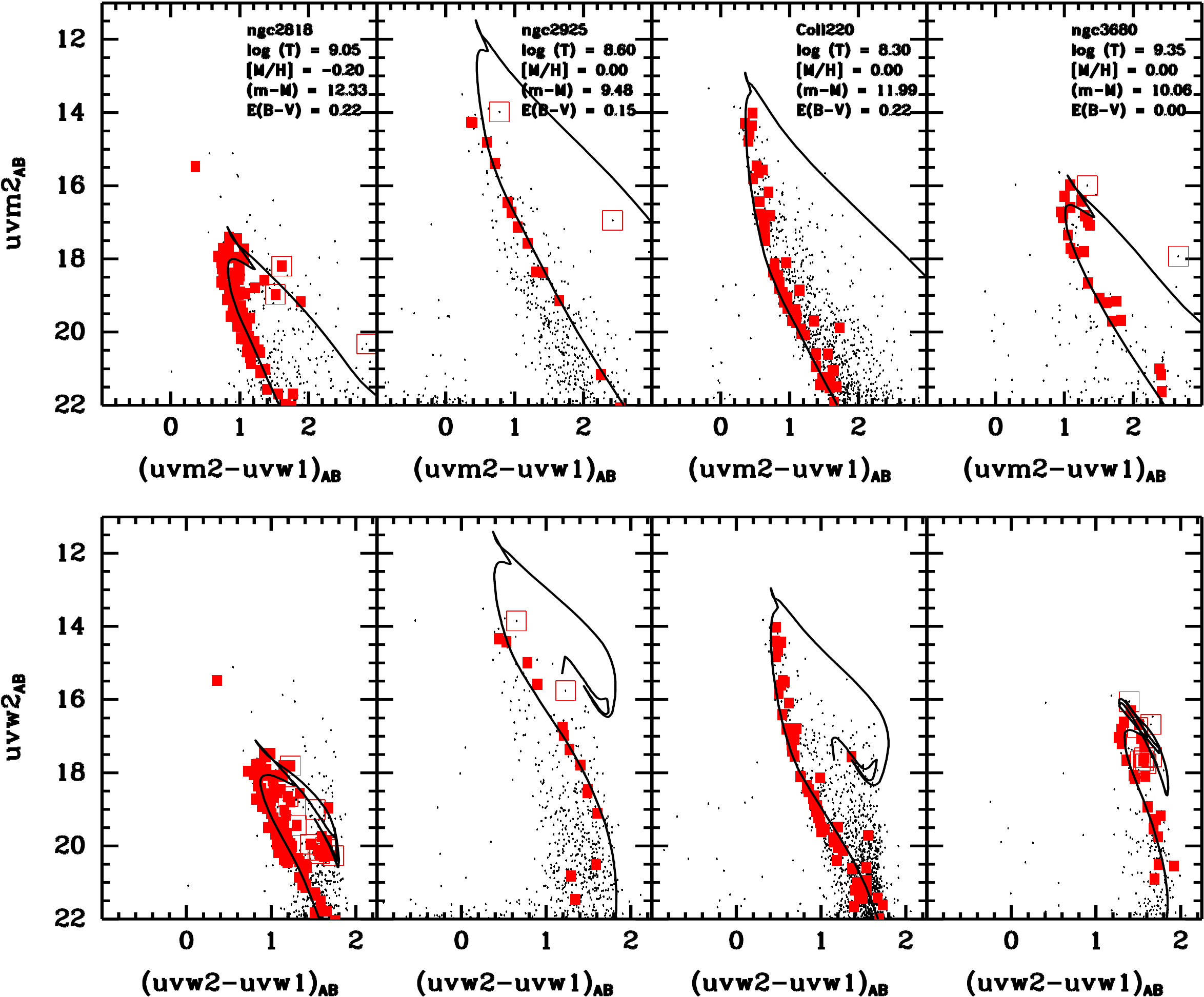}
\end{center}
\caption{Color-magnitude diagrams of the open clusters NGC~2818, NGC~2825, Collinder~220 and NGC~3680 (left to right). The solid lines are PARSEC-COLIBRI isochrones
set to the parameters in the text.  Solid red squares show astrometric members
selected from GAIA DR2 by CG18 while open squares are spectrosopic members from either MMU or sources given in the text.\label{f:CMDfig9}}
\end{figure}

NGC~2818 is a distant third quadrant cluster situated above the Galactic midplane.  MMU has an extensive radial-velocity membership.
This cluster was recently studied in great detail by Bastian et al. (2018) who found that it has a prominent eMSTO. Their complementary
radial velocity study showed that stars on the red side of eMSTO were fast rotators compared to stars on the blue side, confirming the hypothesis
of stellar rotation as the origin of the eMSTO.

The CMDs (Figure \ref{f:CMDfig9}, first column) show a well-defined MS with a prominent MSTO. We find that the cluster
is a bit older than the previous studies, with the MSTO corresponding to an age of approximately 1 Gyr with slightly less-reddening
than Bastiat et al. (2018) of $E(B-V)=0.15$).  One key difference may be that Bastiat et al. assume solar metallicity while we assume a metallicity of [M/H]=-0.2 based
on the analysis of Mermilliod et al.(2001).  Increasing the [M/H] to solar metallicity would not affect the distance or reddening but would
reduce the age of the cluster slightly to 900 Myr.  We identify one member star beyond this turnoff, which could be either be a blue straggler or a disk contaminant.

\begin{figure}[h]
\begin{center}
\includegraphics[scale=.5,angle=0]{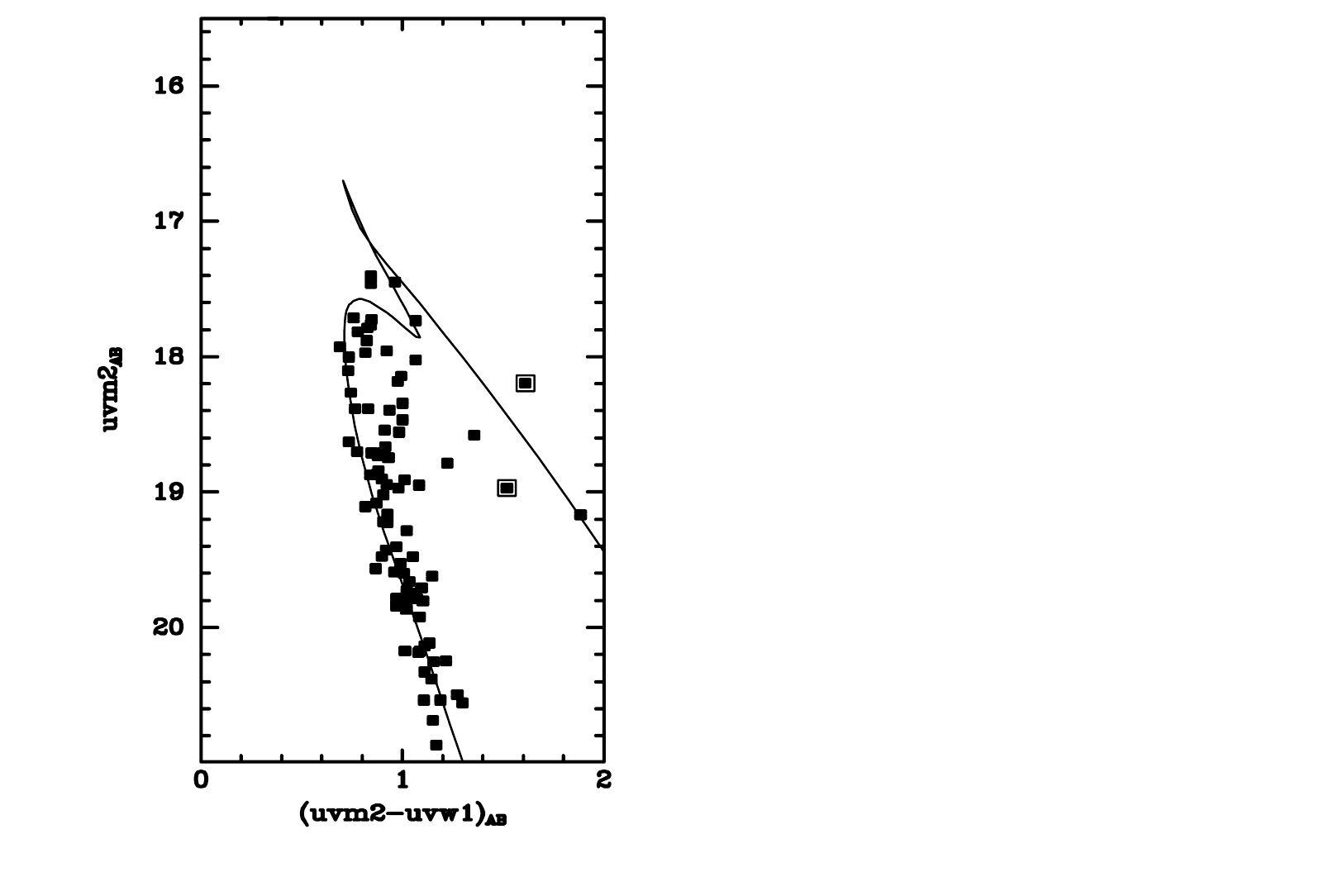}
\end{center}
\caption{A close-up of the MSTO region of NGC~2818.  The left panel shows GAIA- and spectroscopically-selected members. The MSTO region is very broad due to the combination
of eMSTO, convective hook and the beginning of the RGB.  The right panel shows deviations from a linear fit to the MS region and shows a broad pattern in the MSTO
region, again consistent with an eMSTO in NGC~2818 likely resulting from differences in stellar rotation among the MSTO stars.\label{f:NGC2818msto}}
\end{figure}

As noted, Bastian et al. identified an eMSTO in this cluster.  As with NGC~2360, the overlap of convective hook, RGB and MSTO make analysis tricky.  However,
the $uvm2-uvw1$ diagram
shows that the MSTO is broader and more structured (showing what looks like a bifurcated MSTO) than one would expect from a simple stellar population.
Figure \ref{f:NGC2818msto} shows a close-up of this turnoff region and it appears to confirm the broadening of the MSTO.
The right panel shows the deviation from a linear fit to the main sequence to further highlight this feature.

\subsubsection{NGC~2925}

NGC~2925 is a fourth quadrant cluster situated just below the Galactic midplane.  The only focused study is that of Topaktas (1981) who concluded
that the cluster was moderately reddened ($E(B-V)=0.08$) based on photographic photometry.  K16 estimate an intermediate age of 500 Myr.

While the astrometric selection produces a clean sequence in the CMD (Figure \ref{f:CMDfig9}, second column) we note that the three RGB stars cross-identified from MMU lie
outside of the astrometric selection (two of which are detected by UVOT).  In proper-motion space, they span a range of 5 mas yr$^{-1}$, which is a stark contrast
to other clusters in which spectroscopically-confirmed members span at most a few tenths of a mas yr$^{-1}$. Applying our own analysis to the GAIA
data results in an almost identical membership selection to CG18 and confirms that the three MMU members are outliers.
We therefore find it likely that the MMU stars are {\it not} members of NGC~2925 but, given the high proper motions, are halo contaminants.
The astrometric members favor the redder sequence in the field.  Fitting the M17 isochrones produces a fit very similar to K16 but with an increase in reddening to $E(B-V)=0.15$.

\subsubsection{Collinder~220}
\label{ss:coll220}

Collinder~220 is a fourth-quadrant cluster within the Galactic midplane. It has not been the subject of any published individual study and our parameters in Table 2 are taken from K16.
UVOT obtained very deep imaging of this cluster due to its proximity to 1E1024.0-5732 (WR21a), a colliding wind binary near the Westerlund~2 cluster that was intensely monitored
through periastron.

Analysis of this cluster proved challenging.  The VPD shows a clear tight clump of stars within the field star distribution.  However, the field is so dense that our double-Gaussian
modelling failed.  Picking out stars in this clump by hand from within the clump showed (1) a tight sequence of stars in the CMD (Figure \ref{f:CMDfig9}, third column);
(2) a concentration of stars at the nominal position of the cluster; (3) a 
mean proper motion roughly consistent with the K13 parameters ($(\mu_{\alpha}, \mu_{delta}) \sim (-7.3, 2.6)$ compared to (-5.9, 2.5) of K13).  The analysis of CG18 identifies
most of the same stars as members albeit with a more conservative selection. We are therefore confident that this is indeed the cluster sequence.

The derived parameters are markedly different from those of K16. We measure a much longer distance ($m-M=11.99$) and a younger age (200 Myr).
What is really interesting about this cluster, however,
is that it the first to require a change in the reddening law in order to properly fit the color-magnitude sequences.  Most of our clusters were picked to have low foreground extinction
and even those with more extinction are usually well-fit by the standard Milky Way law.  However, using the Milky Way law on Collinder~220 produces a discrepancy in the colors of the
two main sequences which no combination of parameters could fit. Adopting the LMC extinction law, by contrast, results in a consistent fit for both CMDs.  Given the disagreement between our
measured parameters and those of K16, we put forward this interpretation with caution.  However, spectroscopic study or multi-wavelength study of
Collinder~220 would seem warranted to confirm or refute any
changes in the UV extinction law toward the cluster.

\subsubsection{NGC~3680}

NGC~3680 is a fourth quadrant cluster placed slightly above the Galactic midplane. It has been the subject of detailed abundance studies
(Santos et al. 2009; Pena Suarez et al. 2018)
as well as photometric investigations (Anthony-Twarog \& Twarog 2004) which have described it as an old (1.75 Gyr) near-solar metallicity disk cluster.

The astrometric selection produces a clear
main sequence that contains all of the MMU radial velocity members (Figure \ref{f:CMDfig9}, last column).  The isochrones, however, favor a solution
with minimal or no reddening, which produces a consistent fit with a slightly shorter distance ($m-M=10.06$) and
an older age of 2.2 Gyr.

\subsubsection{Lynga~2}

\begin{figure}[h]
\begin{center}
\includegraphics[scale=.5,angle=270]{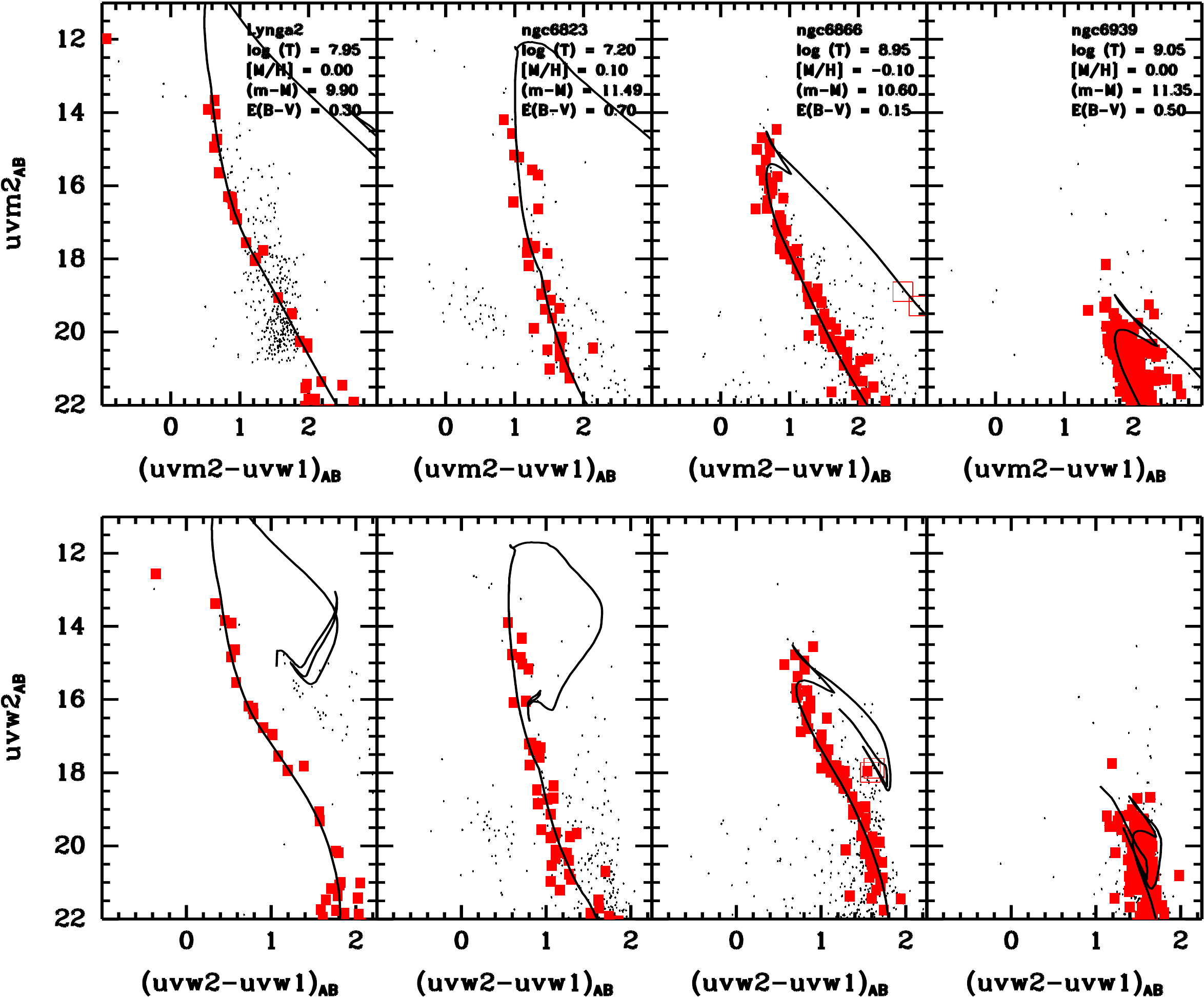}
\end{center}
\caption{Color-magnitude diagrams of the open clusters Lynga~2, NGC~6823, NGC~6866, NGC~6939 (left to right). The solid lines are PARSEC-COLIBRI isochrones
set to the parameters in the text.  Solid red squares show astrometric members
selected from GAIA DR2 by CG18 while open squares are spectrosopic members from either MMU or sources given in the text.\label{f:CMDfig10}}
\end{figure}

Lynga~2 is a small obscured cluster situated toward the Galactic Center. The cluster has been poorly studied, likely due to the high
density of field stars against a relatively spartan cluster.  Alter et al. (1970) and Bica et al. (2006) describe it as young (90 Myr), with moderate
reddening ($E(B-V)=0.22$).

The astrometric and spectroscopic selection produces a spartan but clear MS in the CMDs (Figure \ref{f:CMDfig10}, first column).  We find that an isochrone using the parameters of Bica et al.
follows the color-magnitude sequence quite well (assuming solar metallicity). We note, however, that the brightest member stars are saturated and therefore we are only able to
apply an upper limit to the age of the cluster.

\subsubsection{NGC~6823}

NGC~6823 is a distant obscured first quadrant cluster situated near the midplane.  It is the only very young ($t < $20 Myr) cluster in our program that is well-observed
although it still has a number of saturated stars. It has not been the subject of individual study so parameters are taken from K16.

The astrometric selection produces a clean, if spartan, MS (Figure \ref{f:CMDfig10}, second column). Adjusted
to lower reddening ($E(B-V)=0.70$) and longer distance ($m-M=11.49$) than given in K16, we get excellent agreement
between the isochrone and the sequence.  The MS is a bit broader than the photometric errors, possibly indicating differential reddening or the influence of stellar
rotation.  The brightest stars in this cluster are saturated and so only an upper limit on the age can be given. However, this upper limit is close to the K16 age. Given
the high-quality data, young age and foreground extinction, this cluster would be a good candidate for future spectroscopic study.

\subsubsection{NGC~6866}

NGC~6866 is a first quadrant cluster situated within the Galactic midplane.  Previous studies have shown it to be of intermediate age (900 Myr), moderately reddened
($E(B-V)=0.16$) and slightly metal-poor with $[Fe/H]=-0.10$ (Janes et al. 2014, Bostanci et al. 2015)

The CG18 selection provided to be too conservative -- identified very few clusters members -- and so we applied our own astrometric selection (\S \ref{ss:ngc2343}).  The VPD shows a clear 
clump at the edge of the field star distribution that includes the two MMU radial velocity members and produces a clear MS (Figure \ref{f:CMDfig10}, third column).
Fitting the M17 isochrones to the sequence produces parameters similar to previous investigations.

\subsubsection{NGC~6939}

NGC~6939 is a second quadrant cluster positioned 12 degrees above the Galactic midplane.
Numerous studies have been done of the cluster and have found it to be of solar abundance (Jacobson et al. 2007), heavily reddened ($E(B-V)=0.33$) and old, with
an age of 1.3-1.6 Gyr (Andreuzzi et al. 2004, Rosvick \& Balam 2002). However,
there are significant disagreements between the exact parameters derived from the studies, as shown in Table \ref{t:clusparam}.

The astrometric selection identifies a faint MS in the CMDs (Figure \ref{f:CMDfig10}, last column).  We find that we can fit this sequence using parameters
similar to those of Andreuzzi et al. (2004) but only with some additional reddening ($E(B-V)=0.50$).  This is one of the few clusters in our sample that has substantial reddening
and it is impossible
to line up the main-sequence in both color panels without increasing the reddening.  At this level of extinction, the $uvw2-uvw1$ isochrones show portions of the subgiant
and giant branch to be {\it bluer} than the MSTO because we are only measuring the degree of red leak in these cool and heavily reddened stars.
The CMD also shows at least one star bluer and brighter than the MSTO.  This is a proper-motion member and crudely along the MS that a younger stellar population would follow.  It is likely
that it is a blue straggler.

\subsubsection{NGC~6991}

\begin{figure}[h]
\begin{center}
\includegraphics[scale=.5,angle=270]{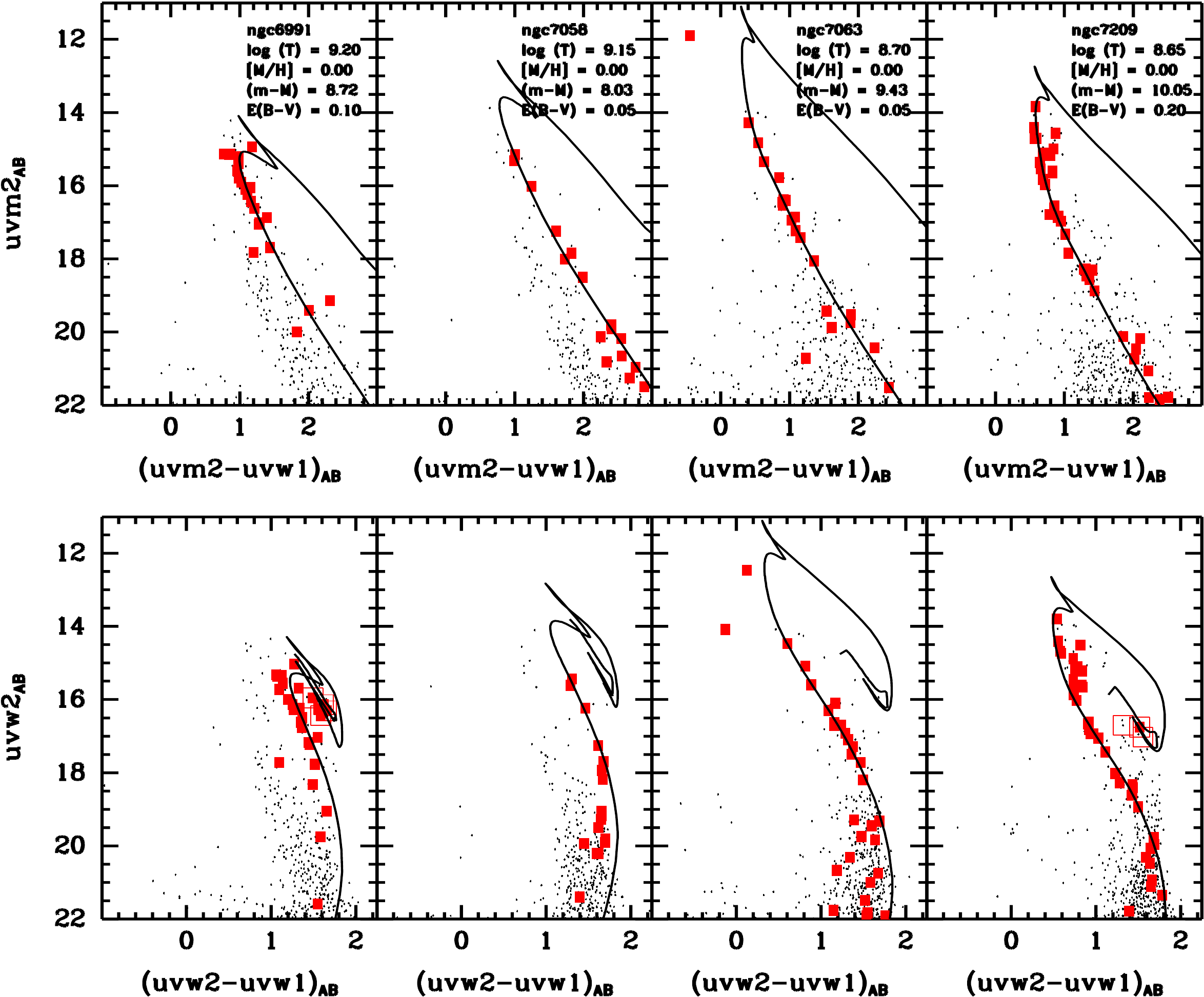}
\end{center}
\caption{Color-magnitude diagrams of the open clusters NGC~6991, NGC~7058, NGC~7063 and NGC~7209 (left to right). The solid lines are PARSEC-COLIBRI isochrones
set to the parameters in the text.  Solid red squares show astrometric members
selected from GAIA DR2 by CG18 while open squares are spectrosopic members from either MMU or sources given in the text.\label{f:CMDfig11}}
\end{figure}

NGC~6991 is a poorly studied first quadrant midplane cluster.  The only estimates of age and distance come from the global survey of K16.
Casamiquela et al.(2016) studied the radial velocity but did not measure metallicity. The cluster is quite large (K13 measure a radius of 26\farcm7) and thus
we only image the central regions.

Th first column of gigure \ref{f:CMDfig11} shows the CMDs of the cluster.  The GAIA astrometry matches a radial velocity member from Casamiquela et al. (2016).  These stars define a narrow
CMD sequence. We measure parameter similar to K16, but with an older age (1.6 Gyr).

\subsubsection{NGC~7058}

NGC~7058 is a nearby sparse second quadrant cluster that is within the Galactic midplane.  It has not been the target of individual study, with the only
estimates of its age, distance and reddening from the global survey of K16.

The CG18 selection selected very few member stars and we therefore applied our method (\S \ref{ss:ngc2343}).
The VPD shows a clear clump of stars well outside of the field star distribution. This clump corresponds to a sequence in the CMD that reveals NGC~7058 to be -- under
the assumption of solar metallicity -- parameters similar to K16 (Figure \ref{f:CMDfig11}, second column). We estimate an age of 1.4 Gyr but this is likely an upper limit.
NGC~7058 is sparse and the images show
a small number of bright saturated stars, at least two of which have the same proper motion as the cluster.  We note that NGC~7058 is another
cluster that shows deviation from the isochrones for faint stars in the $uvw2-uvw1$ color-magnitude diagram.

\subsubsection{NGC~7063}

NGC~7063 is a first quadrant cluster situated nine degrees below the Galactic midplane.  The only comprehensive study to date is that of Pena et al. (2007)
who characterize it as young (140 Myr) and with low foreground reddening ($E(B-V)=0.05$). The MMU survey observed two stars but classified both as non-members.

The CG18 selection proved too conservative and we therefore applied our methods (\S \ref{ss:ngc2343}).
The VPD shows a clear clump of stars outside of the field star distribution. This clump corresponds to a sequence in the CMD that reveals NGC~7063 to be -- under
the assumption of solar metallicity -- roughly consistent with the parameters of Pena et al 2007 (Figure \ref{f:CMDfig11}, third column).  We caution that our estimated age is again an
upper limit, given the presence
of several saturated stars and the curvature of the upper MS.  As with other clusters, NGC~7063 shows deviation from the isochrones for faint stars in the $uvw2-uvw1$ color-magnitude diagram.

\subsubsection{NGC~7209}

NGC~7209 is a second quadrant cluster positioned seven degrees below the Galactic midplane. No spectroscopic metallicity has been published but
photometric studies describe the cluster as solar metallicity, intermediate age (500 Myr) and moderately reddened, with $E(B-V)=0.17$ (Vansevicius et al. 1997, Paunzen et al. 2010, Netopil et al. 2016, K16).

The astrometric and spectroscopic selection find a clear MS in the CMDs (Figure \ref{f:CMDfig11}, last column).  Fitting the M17 isochrones to this sequence yields parameters similar to the literature.
However, we note that the CMD shows significant structure that almost resembles a second turnoff.  Increasing the age of the cluster to 700 Myr would match this second turnoff albeit
with some slight discrepancy in the curvature of the upper MS.  This would relegate the brighter stars to being blue stragglers. If the brighter stars represent the MSTO, then NGC~7209 would
have an age of 500 Myr but an eMSTO.  Spectroscopic investigation of the rotation rates could determine if this is the case.


\begin{deluxetable}{cccccc}
\tablewidth{0 pt}
\tabletypesize{\footnotesize}
\tablecaption{Parameters of Clusters in this Study\label{t:clusparam}}
\tablehead{
\colhead{Cluster} &
\colhead{Distance} &
\colhead{E (B-V)} &
\colhead{log(Age)} &
\colhead{[Fe/H]} &
\colhead{Source}\\
\colhead{} &
\colhead{m-M} &
\colhead{} &
\colhead{(yr)} &
\colhead{} &
\colhead{}}
\startdata
         NGC~752     &  8.30  &  0.03  &    9.15  & -0.03   &  B{\"o}cek Topcu et al. (2015), Twarog et al. (2015)\\
         NGC~752     &  8.30  &  0.05  &    9.15  & +0.00   &  This Study \\
\hline
         NGC~1039    &  8.38  &  0.10  &    8.35  & +0.07   &  Jones \& Prosser (1996) \\
         NGC~1039    &  8.71  &  0.07  &    8.25  & +0.07   &  Schuler et al. (2003) \\
         NGC~1039    &  8.54  &  0.08  &    8.38  &  0.07   &  Kharchenko et al. (2013) \\
	 NGC~1039    &  8.75  &  0.10  & $<$8.20  & +0.10   &  This Study\\
\hline
	NGC~2192     & 12.70  &  0.20  &    9.04  & -0.31   &  Park \& Lee (1999)\\
	NGC~2192     & 12.70  &  0.16  &    9.15  & -0.31   &  Tapia et al (2010)\\
	NGC~2192     & 12.70  &  0.15  &    9.20  & -0.30   &  This Study\\
\hline
	NGC~2204     & 13.00  &  0.13  &    9.20  & -0.23   &  Kassis et al. (1997), Jacobson et al. (2011)\\
	NGC~2204     & 13.16  &  0.00  &    9.35  & -0.20   &  This Study\\
\hline
	NGC~2243     & 12.78  &  0.11  &    9.67  & -0.42   &  Twarog et al. (1997), Jacobson et al. (2011)\\
	NGC~2243     & 12.91  &  0.00  &    9.70  & -0.50   &  This Study\\
\hline
	NGC~2251     & 10.60  &  0.20  &    8.50  & -0.10   &  Parisi et al. (2005), Reddy et al. (2013)\\
	NGC~2251     & 10.61  &  0.25  &    8.55  & -0.10   &  This Study\\
\hline
	NGC~2281     &  8.55  &  0.09  &    8.80  &  0.00   & Glaspey (1987), Netopil(2017)\\
	NGC~2281     &  8.56  &  0.15  &    8.80  &  0.00   & This Study\\
\hline
\hline
\enddata
\tablenotetext{a}{Using LMC extinction law.}
\footnotesize{This table is available in its entirety in machine-readable and Virtual Observatory (VO) forms.}
\end{deluxetable}

\section{Discussion}
\label{s:discussion}

\subsection{Isochrones and Cluster Fits}

The purpose of this program was to test the utility of standard isochrones in the UV.  In that sense, we find that the isochrones perform quite well,
successfully reproducing the upper main sequence for cluster up to several Gyr in age.  Most of the CMDs are well-described
using isochrones set to values at or near those in the literature.  For clusters with significant revisions, the source of the discrepancy is either
older studies or, in some cases, previous studies mistaking the disk sequence for the cluster sequence.

The consistency between the models and the data shows that even for fairly young stellar populations, the NUV emission is well-described by the existing
atmospheric models.  There is little evidence of stochastic variation and little evidence of the excess emission produced by chromospheric activity
as seen in the FUV (see discussion in Smith 2018). However, it is important to note that our study is focused on only  limited range of parameter space.
Almost all of our clusters are older than
100 Myr and the brightest stars in the youngest are few and saturated.  Younger clusters will host much more massive stars, in which factors 
like binarism, rotation and magnetic field can play a more important
role and our understanding of convention and mixing are poor (see, e.g., Viallet et al. 2013).
In addition, our stars are primarily metal-rich (or assumed to be) so we are unable to test how
the isochrones perform at lower metallicity levels.  Finally, almost all of our clusters are minimally reddened with only a handful having significant extinction.
Unfortunately, these limitations can not really be exceeded with the Milky Way's contingent of open cluster or with UVOT.  We are currently investigating
the UV properties of old metal-poor populations in globular clusters. But exploring very young ($< 100 Myr$) stellar populations or populations with significant
reddening would require surveys of the nearby stellar populations in the Magellanic Clouds.  Swift/UVOT has also surveyed these areas (Hagen et al., 2017) but its
reach may exceed its grasp in such dense star-forming regions as 30 Doradus where the younger stars tend to reside. That is more properly the domain of future
missions that have the wide-field high-resolution multi-filter FUV imaging.

The main deficiencies in the isochrones are for redder stellar types: the RGB, AGB and lower MS stars that are cool and have little emission in the UV.
This is hardly a surprise given that for cool stars (those with $T_{eff} < 7000 K$), the ``UV" signal in two of Swift's filters becomes dominated by the red leak.
This plays havoc with the effective wavelength (see, e.g., Siegel et al. (2012)) and causes the reddening to have a non-linear effect (Paper I, Brown et al. 2010). For
later stellar types that are heavily extinguished, this will cause them to move {\it blueward} in $uvw2-uvw1$ CMDs as the $uvw1$ ramps up the red leak faster than the
$uvw2$ filter.

However, it could also be produced by deficiencies in either the isochrones or the underlying atmospheric models.  Such discrepancies have been noted before
in NUV photometry of cool stars (see, e.g., Barker \& Paust 2018) and are likely the result of the underlying atmospheric models having an incomplete treatment of UV absorption lines
for very cool atmospheres.  This problem has been discussed in detail in the context of the UVBLUE project by Rodr{\'{\i}}guez-Merino et al.(2005, 2009)
and Chavez (2007) and is part of the motivatio for the {\it Hubble Space Telescope}'s upcoming Hubble UV Legacy Library of Young Stars As Essential
Standards (ULLYSES) program.

\begin{figure}[h]
\begin{center}
\includegraphics[scale=.7,angle=0]{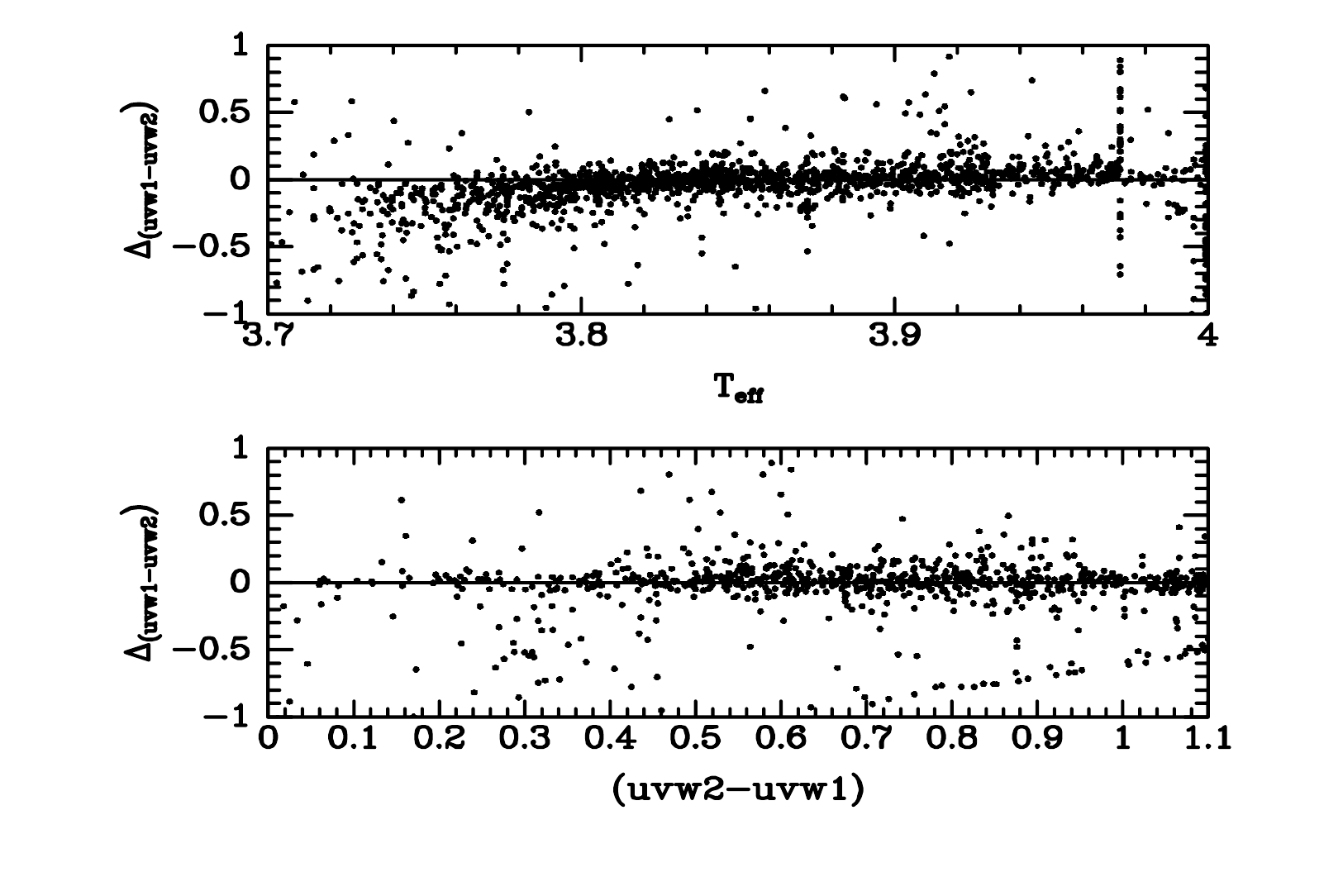}
\end{center}
\caption{A comparison of photoemtric distance from the fit isochrones for a sample of the UVOT cluster stars, plotted against absolute $T_{eff}$ (bottom)
and observationsl $uvw2-uvw1$ color (top). The solid line shows agreement.  The vertical streak in the top panel is NGC~2658, where the MS is cut off slightly
fainter than the blue straggler sequence. 
\label{f:isocheck}}
\end{figure}

To determine the nature of these discrepancies, we examined the deviations of stellar measures from the fit isochrones. This deviation was calculated
by measuring, for each star, the minimum quadrature color-magnitude distance from the isochrone.  Figure \ref{f:isocheck} shows the residuals plotted
against observational $uvw2-uvw1$ color and the effective temperature implied by the isochrone.
If the problem arises from the red leak, it should
occur at a consistent {\it observed} color, where the red leaks takes over the signal. However, it the problems arises from atmospheric models, it should occur at a
consistent effective temperature, where the atmospheric models begin to fail.  For clusters that show the deviation at faint magnitudes,
the discrepancies do not track observed stellar color very well.  The deviations do, however,
track effective temperature quite well, beginning to deviate toward bluer colors at approximately $log T_{eff}\sim3.8$ or 6300 K.
This hints that the problem lies
within the stellar atmopshere models. However, for the bulk of stars that are of interest
in the UV, the atmospheric models and isochrones hold up quite well.

\subsection{Reddening Law}

As noted above, most of our program clusters have low intrinsic reddening.  Those few that have substantial reddening seem to work fine with a ``Milky-Way"-like extinction
law (i.e., one with a shallower slope and a strong 2175 \AA\ bump).  Only one cluster -- Collinder~220 -- shows any sign of deviation from the standard reddening law. This is somewhat
surprising as previous investigations have shown some variation in the extinction law at high latitudes within the Galaxy (Peek \& Schiminovich 2013), at
low latitude within the Galaxy (Siegel et al. 2012) and in nearby galaxies (Hagen et al. 2017, 2019).  This suggests that
future endeavors should both look at resolved stellar population in nearby galaxies where the reddening law is known to vary and at more reddened clusters within
our own Galaxy where variations in the law will be amplified and more readily detectable.  It might also be of benefit to make these studies multi-wavelength, as in 
Hagen et al. (2017, 2019) where the more well-understood OIR extinction law can be used to ``nail down" the line-of-sight extinction, freeing the UV to focus more on
variation in the law instead of variation in the total extinction level along the line of sight.

\subsection{Extended MSTOs}

Two of our clusters -- NGC~2360 and NGC~2818 -- have previously been identified as having extended MSTOs based on photometric and spectroscopic exploration. We confirm
the photometric broadening and further note that a number of other clusters (NGC~2355, NGC~2420 NGC~2658, NGC~6823, NGC~7209) show a similar broadening of the MSTO beyond
what is expected from photometric error. This is unlikely to be the effect of
differential reddening because most of our clusters were selected to have low reddening and differential reddening would broaden the entire MS. Spectroscopic investigation of
rotation rates would confirm the nature of these extended MSTOs, as it has for NGC~2360 and NGC~2818.
If confirmed, this would support the contention of Cordoni et al. (2018) that the feature is not unusual among Galactic open clusters and the UV may be more sensitive
to the effect than the optical or IR.

Considering that the eMSTO involves the brightest stars in any specific population, this is an effect that may need to be accounted for in spectral synthesis models. 
While the precise effect of stellar rotation on integrated light is beyond the scope of this paper, the indications that stellar rotation can increase both mass loss rates
and main sequence lifetime makes their potential ubiquity in young stellar populations both critical to confirm and critical to incorporate into the models.

\section{Conclusions}
\label{s:conc}

We have measured photometry of stars in 103 open clusters using the Swift/UVOT telescope.  We have analyzed 49 of these clusters, using GAIA DR2 and spectroscopic studies to separate
members stars from the field to study precise CMDS and compared the CMDs to theoretical isochrones.  Our main results are:

$\bullet$ The theoretical isochrones reproduce the features of the CMDs very well.  The only consistent discrepancy is at faint magnitudes where cool late-type stars are bluer in $uvw2-uvw1$ space
than anticipated. This could be a result of red leak or of inadequacies in the UV opacities in model atmospheres for late-type stars. The discrepancies seem to be more closely related
to absolute magnitude than color, favoring a problem with the model atmospheres.

$\bullet$ Using these isochrones, we measure age, reddening and distance for 49 clusters with well-defined color-magnitude sequences.
For those that have been studied in detail before, we generally find agreement with literature values.  However, there are a number of clusters for which we find significantly
different values than the literature,
particularly in the third Galactic quadrant where some previous studies have mistaken the disk sequence for the cluster.  We catalog substantially revised parameters for the clusters
NGC~2304,  NGC~2343, NGC~2360, NGC~2396, NGC~2428, NGC~2509, NGC~2533, NGC~2571, NGC~2818, Collinder~220 and
NGC~6939.

$\bullet$ We confirm the presence of an extended MSTO in two previously studied clusters -- NGC~2360 and NGC~2818.  We also identify broad MSTO features in at least five other clusters that could
warrant further spectroscopic investigation.

$\bullet$ Most of our clusters have minimal reddening and are thus unsuited to probe the UV properties of the foreground dust.  However, one cluster -- Collinder 220 -- shows significant
improvement in the isochrone fits if an ``LMC-like" reddening law -- one with a smaller red bump at 2175 \AA\ and a steeper extinction curve -- is used.

Our investigation has only scratched the surface of what this remarkable data set can yield.  Measuring the integrated light would allow direct comparison between isochrones
and synthetic spectral models to test the validity of the latter for unresolved stellar populations. Combining this database with
extra ground-based optical photometry or the
broad-band photometry of GAIA DR2 would allow the detection of unresolved white dwarf-main sequence binaries, providing additional insight into
the role of binarism in spectral synthesis models and the integrated luminosity of stellar population (see, e.g., Buzzoni et al. 2012,
Hern{\'a}ndez-P{\'e}rez \& Bruzual 2013).  A more thorough examination of the MSTOs -- especially in combination with spectroscopic
study -- would allow a much more detailed exploration of eMSTOs and their connection to cluster age.

\acknowledgements

The authors acknowledge sponsorship at PSU by NASA contract NAS5-00136.  This research was also supported by the NASA ADAP through grants
NNX13AI39G and NNX12AE28G. The authors thank L. Girardi for useful discussions about the NUV isochrones.
The Institute for Gravitation and the Cosmos is supported by the Eberly College of Science and the Office of the Senior Vice President for Research at the Pennsylvania State University.

\bibliographystyle{apj}

\end{document}